\documentclass[%
reprint,
superscriptaddress,
nobibnotes,
amsmath,amssymb,
aps,
prl,
floatfix,
]{revtex4-2}

\usepackage{graphicx}
\usepackage{dcolumn}
\usepackage{bm}
\usepackage{cancel}
\usepackage{float}
\usepackage[english]{babel}
\usepackage{hyperref}
\hypersetup{
    colorlinks=true,
    linkcolor=blue,
    citecolor=blue,
    urlcolor=blue,
    breaklinks=true
}
\begin{document}

\title{Three-dimensional description of vibration-assisted electron knock-on damage}

\author{Alexandru Chirita}
\author{Alexander Markevich}
\affiliation{University of Vienna, Faculty of Physics, Boltzmanngasse 5, 1090 Vienna, Austria}%
\author{Mukesh Tripathi}
\altaffiliation[Present address: ]{{\'E}cole polytechnique f{\'e}d{\'e}rale de Lausanne (EPFL), BM 2143, Station 17, CH-1015 Lausanne, Switzerland}
\affiliation{University of Vienna, Faculty of Physics, Boltzmanngasse 5, 1090 Vienna, Austria}%
\author{Nicholas A. Pike}
\affiliation{nanomat/Q-MAT/CESAM and European Theoretical Spectroscopy Facility, Universit\'e de Li\`ege, B-4000 Li\`ege, Belgium}
\author{Matthieu J. Verstraete}
\affiliation{nanomat/Q-MAT/CESAM and European Theoretical Spectroscopy Facility, Universit\'e de Li\`ege, B-4000 Li\`ege, Belgium}
\author{Jani Kotakoski}
\author{Toma Susi}%
\email{toma.susi@univie.ac.at}
\affiliation{University of Vienna, Faculty of Physics, Boltzmanngasse 5, 1090 Vienna, Austria}%

\date{November 18, 2021}

\begin{abstract}
Elastic knock-on is the main electron irradiation damage mechanism in metals including graphene. Atomic vibrations influence its cross-section, but only the out-of-plane direction has been considered so far in the literature. Here, we present a full three-dimensional theory of knock-on damage including the effect of temperature and vibrations to describe ejection into arbitrary directions. We thus establish a general quantitative description of electron irradiation effects through elastic scattering. Applying our methodology to in-plane jumps of pyridinic nitrogen atoms, we show their observed rates imply much stronger inelastic effects than in pristine graphene.
\end{abstract}
\maketitle

Transmission electron microscopy (TEM) is a powerful probe of the atomic structure of materials. Due to efficient aberration correction~\cite{hawkes_aberration_2009}, the information that modern instruments can collect is mainly limited by radiation damage~\cite{egerton_radiation_2019}. Knock-on displacement due to electron backscattering~\cite{banhart_irradiation_1999} affects all materials, and is the primary damage mechanism for metals such as graphene~\cite{meyer_direct_2008}. For non-destructive imaging, the electron should not transfer more energy than the displacement threshold of the material, defined as the energy needed to remove an atom at rest from the lattice. Conversely, one may want to activate certain beam-induced processes, where the energy received by the nucleus drives the resulting dynamics~\cite{kotakoski_stone-wales-type_2011, susi_siliconcarbon_2014}, enabling atomically precise manipulation~\cite{susi_manipulating_2017,dyck_placing_2017,tripathi_electron-beam_2018}.

Since the electron mass is small compared to the nuclei, momentum conservation strictly limits the amount of kinetic energy it can transfer in an elastic electron-nucleus collision. However, at low electron energies, the thermal motion of the atom can significantly increase the energy transfer~\cite{susi_quantifying_2019} -- indeed, irradiation damage of graphene at energies below 100~keV cannot be explained without vibrations~\cite{meyer_accurate_2012}. To date, these have only been treated in the out-of-plane direction, where the momentum transfer is most efficient. For a complete description of situations including the beam-induced movement of adatoms~\cite{egerton_beam-induced_2013}, momentum transfers in all directions must be included. Further, while knock-on damage in pristine graphene can be accurately described from first principles~\cite{susi_isotope_2016,chirita_mihaila_influence_2019}, there are puzzling discrepancies between the predicted and measured cross-sections for its impurity sites~\cite{susi_towards_2017}. Until now, it has not been clear if these are due to shortcomings in the elastic model, or arise from unaccounted inelastic effects.

We present a full three-dimensional theory of electron knock-on damage for arbitrarily moving target atoms, and explore its implications for knock-on displacements from pristine graphene as well as for the reversible transformations of its pyridinic nitrogen impurity sites~\cite{lin_structural_2015}, for which we include new quantitative measurements. Such high-quality data provides a rigorous test, but our theoretical framework is generally valid for the quantitative description of elastic electron irradiation effects.

\begin{figure}[ht!]
\centering
    \includegraphics[width=1.0\linewidth]{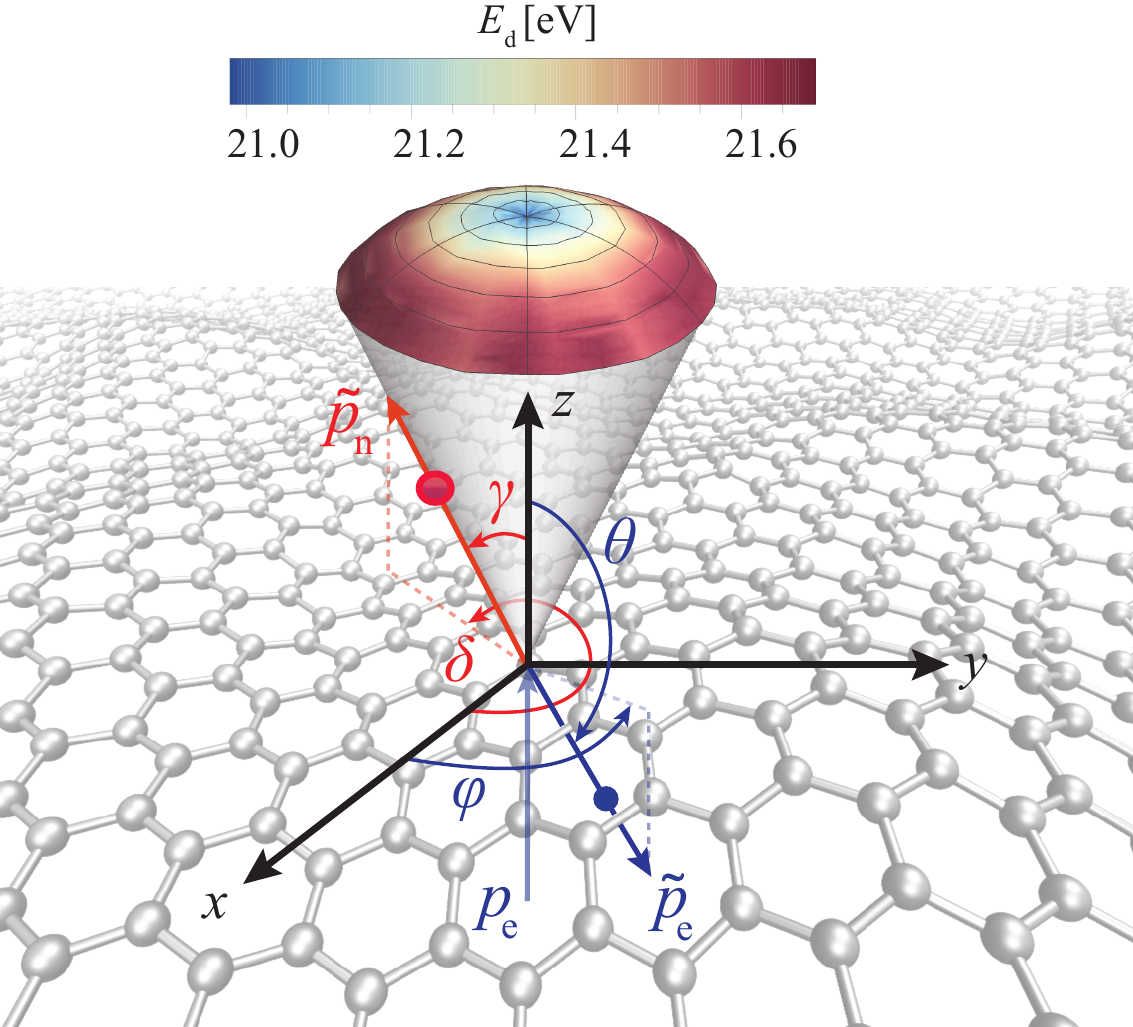}
    \caption{(Color online) The three-dimensional electron-atom scattering geometry. $\varphi$ and $\theta$ are the azimuthal and polar electron scattering angles after the collision (blue), whereas $\delta$ and $\gamma$ are the azimuthal and polar emission angles of the atom (red). The colored dome indicates the calculated angular variation of the displacement threshold energy $E_\mathrm{d}$ for pristine graphene (multiplied by $\xi$=~0.931 to match the experimental cross section, see Table~\ref{table:fitting}).}
    \label{fig:geometry}
\end{figure}

To construct a general model, we need to consider the full scattering process of an electron by an atom. The electron (mass $m$, energy $E_\mathrm{e}$, momentum $p_\mathrm{e}$) is a relativistic projectile that interacts with a moving, non-relativistic target, the nucleus (mass $M$, energy $E_\mathrm{n}$, momentum $p_\mathrm{n}$). Assuming that the electron has an initial momentum along the $z$-axis, $\mathbf{p}_\mathrm{e}=|\mathbf{p}_\mathrm{e}|\left(0,0,1\right)$ with $|\mathbf{p}_\mathrm{e}|=p_\mathrm{e}=\sqrt{E_\mathrm{e}\left(E_\mathrm{e}+2mc^2\right)/c^2}$ ($c$ is the speed of light), and that the targeted atom has initial momentum components in all three Cartesian directions $\mathbf{p}_\mathrm{n}=M\left(v_\mathrm{x},v_\mathrm{y},v_\mathrm{z}\right)$, where $v_{x,y,z}$ define the initial velocity of the nucleus, we derive a three-dimensional description of the electron-nucleus scattering  (Fig.~\ref{fig:geometry} illustrates the geometry). The electron can scatter into any angle after the collision, $\mathbf{\tilde{p}}_\mathrm{e}=|\mathbf{\tilde{p}}_\mathrm{e}|\left(\sin{\theta}\cos{\varphi},\sin{\theta}\sin{\varphi},\cos{\theta}\right)$, and so can the atom, $\mathbf{\tilde{p}}_\mathrm{n}=|\mathbf{\tilde{p}}_\mathrm{n}|\left(\sin{\gamma}\cos{\delta},\sin{\gamma}\sin{\delta},\cos{\gamma}\right)$, where $\varphi$ and $\theta$ are the electron azimuthal and polar scattering angles and $\delta$ and $\gamma$ are the corresponding atom emission angles.

We are interested in the energy transferred to the nucleus $\tilde{E}_\mathrm{n}$, which we can derive from the relativistic energy and momentum conservation (this analytic solution assumes that $E_e \gg \tilde{E}_\mathrm{n} - E_\mathrm{n}$, see the Supplemental Material~\cite{supplement} for details):

\begin{widetext}
\begin{equation}\label{eq:emax3Dtransfer}
\tilde{E}_\mathrm{n}\left(E_\mathrm{e},v_\mathrm{x,y,z},\theta, \varphi \right) =
\frac{M\left(v_\mathrm{x}^2+v_\mathrm{y}^2+v_\mathrm{z}^2\right)}{2} + p_\mathrm{e}\left(1-\cos{\theta}\right)\left[\frac{ p_\mathrm{e}}{M}+v_\mathrm{z}-\sin{\theta}\left(v_\mathrm{x}\cos{\varphi}+v_\mathrm{y}\sin{\varphi}\right)\right].
\end{equation}
\end{widetext}

Note that although electron scattering from the Coulomb potential of the nucleus is spherically symmetric, the inclusion of arbitrary velocity components breaks that symmetry for atom emission. While Eq.~\ref{eq:emax3Dtransfer} provides the general expression of energy transfer in terms of the electron scattering angles $\varphi$ and $\theta$, a connection to the atom emission angles $\gamma$ and $\delta$ is needed to describe its displacement after the collision, which is what can be directly simulated~\cite{kotakoski_stone-wales-type_2011,susi_atomistic_2012}. These emission angles can be derived from momentum conservation (\cite{supplement}; Fig.~\ref{fig:geometry} again shows the geometry); note that $\gamma$ and $\delta$ now depend on the instantaneous velocity of the atom at the moment of the scattering, and yield unique values for any specific initial state:

\begin{widetext}
\begin{eqnarray}
    \gamma\left( E_\mathrm{e},v_\mathrm{x,y,z},\theta,\varphi \right)
    &=& \arctan\left(\frac{\sqrt{\left(Mv_\mathrm{x}-p_\mathrm{e}\,\sin{\theta}\cos{\varphi} \right)^2 + \left(Mv_\mathrm{y}-p_\mathrm{e}\,\sin{\theta}\sin{\varphi}\right)^2}}
    {Mv_\mathrm{z}+p_\mathrm{e}\,\left( 1-\cos{\theta}\right) }\right),\label{eq:gammaMain}
\end{eqnarray}

\begin{eqnarray}\label{eq:deltaMain}
  \delta\left( E_\mathrm{e},v_\mathrm{x,y,z},\theta,\varphi \right)
  &=& \arctan{\left(\frac{Mv_\mathrm{y} -p_\mathrm{e}\,\sin{\theta}\sin{\varphi}}{Mv_\mathrm{x}-p_\mathrm{e}\,\sin{\theta}\cos{\varphi}}\right)}.
\end{eqnarray}
\end{widetext}

To describe the knock-on process for pristine graphene, we ran a series of density functional theory molecular dynamics (DFT/MD) simulations using our established methodology~\cite{susi_siliconcarbon_2014,susi_isotope_2016} for atom emission angles between $0^\circ\leq \gamma \leq 30^\circ$ (as reduced energy transfer for larger angles cannot overcome the displacement threshold \cite{supplement}) and $0^\circ \leq \delta \leq 60^\circ $, which by symmetry allows us to predict $E_\mathrm{d}\left(\gamma,\delta\right)$ for all azimuthal angles. We used the Atomic Simulation Environment~\cite{larsen_atomic_2017} for Velocity-Verlet dynamics with a timestep of 0.3\,fs on a 7$\times$7$\times$1 graphene supercell, with forces from a GPAW~\cite{enkovaara_electronic_2010} DFT calculator using the PBE functional~\cite{perdew_generalized_1996}, a \textit{dzp} basis set, a 3$\times$3$\times$1 Monkhorst-Pack $\mathbf{k}$-point grid, and a Fermi-Dirac smearing of 0.025\,eV (which results in notably higher values of $E_\mathrm{d}$ than the default setting of 0.1\,eV). $E_\mathrm{d}$ increases for emission angles $\gamma > 0^\circ$, as shown in Fig.~\ref{fig:geometry} (and noted earlier \cite{zobelli_electron_2007}). Combined with the fact that a normally incident electron can transfer the less energy the larger the angle $\gamma$ is, this illustrates why the out-of-plane approximation has worked so well for this typical geometry.

In experimental studies, only the rate or probability of displacements can be measured. Electrons displace atoms via Coulomb interaction with the nuclei, and the relativistic scattering cross-section between an electron and a target atom has been derived by Mott~\cite{mott_theory_1965} and expanded by McKinley and Feshbach~\cite{mckinley_coulomb_1948} to obtain an analytical expression $\sigma(E_e,\theta)$ \cite{supplement}, accurate up to middle-$Z$ elements. Defect creation in pristine graphene starts a few tens of keV below the static $E_d$, explained by accounting for the vibrational enhancement of the momentum transfer~\cite{meyer_accurate_2012}. Most calculations have assumed that $E_\mathrm{d}$ is isotropic~\cite{seitz_notitle_1956}, and that only out-of-plane vibrations are important~\cite{susi_isotope_2016} (if any). We expand this to a fully three-dimensional model including vibrations and momentum transfers in an arbitrary direction, giving the cross section $\sigma_\mathrm{3D}$ as an integral over the electron scattering angles and the nuclear velocity components~\cite{supplement}:

\begin{widetext}
\begin{equation}\label{eq:sigma3d}
\sigma_\mathrm{3D}(E_\mathrm{e}, E_\mathrm{d}, T)= \int_0^{2\pi}\int_0^{\pi}\prod_{i=x,y,z}\int\displaylimits_{-v_i^\mathrm{max}}^{v_i^\mathrm{max}}
P(v_i,\overline{v_i^2}(T))\,\mathcal{H}(\tilde{E}_\mathrm{n}-\xi {E_\mathrm{d}}\left(\gamma,\delta\right) )\,\sigma(E_\mathrm{e},\theta)\,\mathrm{d}v_i\,\sin{\theta}\,\mathrm{d}\theta\,\mathrm{d}\varphi,
\end{equation}
\end{widetext}
where 
$P(v_i,\overline{v_i^2}(T))$ with $i={x,y,z}$ are the normal distributions of Cartesian atom velocities with mean-square widths $\overline{v_i^2}(T)$ derived from the phonon density of states and integrated over $\pm v_i^{max}$ covering their variation~\cite{susi_isotope_2016}, and $\mathcal{H}$ is a Heaviside step function ensuring the transferred energy $\tilde{E}_\mathrm{n}$ (Eq.~\ref{eq:emax3Dtransfer}) exceeds the value of the angle-dependent displacement threshold ${E_\mathrm{d}}\left(\gamma,\delta\right)$ \cite{supplement}. The latter is multiplied by a fitting factor $\xi$ to match the experimental cross-section. This demanding five-dimensional expression could only be integrated using adaptive numerical methods \cite{supplement}.

\begin{figure}[ht!]
\centering
    \includegraphics[width=0.97\linewidth]{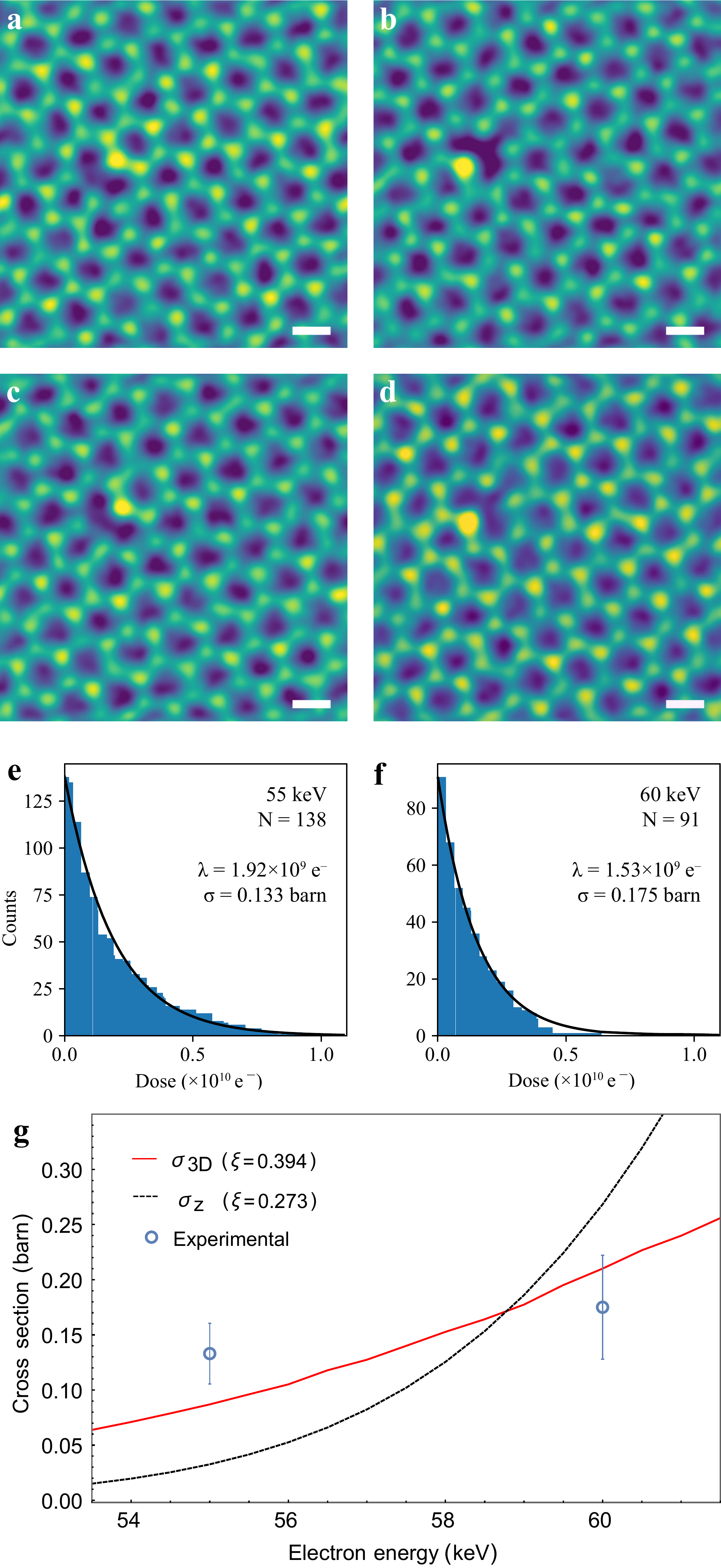}
    \caption{(Color online) Reversible jumping of the N--C$_2$ impurity across a vacancy in graphene. (a--d) Four frames from an annular dark field scanning transmission electron microscopy image series recorded at 60~keV at room temperature. The scale bar is 2\,\AA. (e, f) Exponential distribution of jump doses ($N$ in total) fitted with the expected Poisson $\lambda$ dose at (e) 55 (resulting cross section $\sigma$ of 0.133\,barn) and (f) 60\,keV (0.175\,barn). (g) Comparison of calculated cross sections with $z$-velocity only (black dashed line) and the full 3D model (red line) with the measurements. (The subtle undulation of the 3D curve is due to numerical variation.)}
    \label{fig:pyrjumping}
\end{figure}

We further considered the pyridinic N site (N--C$_2$), which is similar to a single vacancy, but with the dangling-bond C atom replaced by N, bonding to two C neighbors and stabilizing the defect. These N have been observed to rapidly ``jump" back and forth across the vacancy under electron irradiation at 500\,$^\circ$C~\cite{lin_structural_2015}. The calculated energy barrier of $\sim$4~eV for the process is too high to be thermally activated, but neither was it possible to explain the observed event rate with the earlier models~\cite{susi_towards_2017}. Since the predicted minimum-energy pathway is in-plane, $x,y$ velocities presumably play a crucial role in activating the jumps. We collected new room-temperature data at 55 and 60\,keV on incidental impurities identified by electron energy loss spectroscopy and obtaining precise cross-section estimates (see Fig.~\ref{fig:pyrjumping}).

We ran DFT/MD for various emission angles at the N--C$_2$ site to determine under which conditions the N can cross the vacancy to bind with the C atoms on the other side. Using phonon modeling
via first-principle calculations~\cite{supplement} following our earlier methodology~\cite{tripathi_electron-beam_2018}, we estimated mean-square velocities of \{$\overline{v^2_x}=4.81$, $\overline{v^2_y}=3.72$, and $\overline{v^2_z}=2.14\}\times10^5$\,$\mathrm{m}^2\mathrm{s}^{-2}$ for the N atom (where the $y$ direction points across the vacancy) to quantify the effect of its in- and out-of-plane motion.

Maximum energy transfers near electron back-scattering do not result in a jump, regardless of the initial momentum of the N atom. Instead, effective emission directions lie in a narrow sector between $90^\circ \leq \delta \leq 110^\circ$ (and symmetrically towards the other side, where $\delta = $ 90$^\circ$ points directly across the vacancy) and $55^\circ \leq \gamma \leq 90^\circ$, with $E^{\mathrm{N}}_\mathrm{d}\left(\gamma,\delta\right)$ ranging from 10 to 13.5\,eV ~\cite{supplement}. We interpolated the resulting threshold values and calculated cross-sections with Eq.~\ref{eq:sigma3d}. However, we had to drastically scale down $E_\mathrm{d}\left(\gamma,\delta\right)$ with the multiplicative factor $\xi$ (see $\sigma_\mathrm{3D}$ in Table~\ref{table:fitting}) to match the experiment. Nonetheless, it is clear from Fig.~\ref{fig:pyrjumping}g that the 3D model describes the experimental data much better.

Finally, we explored including the variation of $E_\mathrm{d}$ due to thermal perturbations from the equilibrium geometry~\cite{chirita_mihaila_influence_2019}. Unfortunately, the addition of an energy dimension to the numerical integration of Eq.~\ref{eq:sigma3d} proved too demanding, nor is it possible to perform the required thermal sampling at our DFT/MD level of theory. However, this effect is expected to be only a minor correction~\cite{chirita_mihaila_influence_2019}.

\begin{table*}[t!]
\caption{Different cross-section models fitting experimental data for carbon displacement from $^{12}$C graphene and the pyridinic N jump (N--C$_2$). $\sigma_\mathrm{z}$ is the $v_z$-dependent cross-section~\cite{susi_isotope_2016,chirita_mihaila_influence_2019} and $\sigma_\mathrm{3D}$ is the cross-section with the non-isotropic threshold energy $E_\mathrm{d}\left(\gamma,\delta\right)$ cross-section. $\xi$ is a fitting factor for the simulated displacement threshold energies $E_\mathrm{d}\left(\gamma,\delta\right)$ used to match the experimental displacement cross-sections, and WMSE is the weighted mean squared error used for the fit. Only the lowest $E_\mathrm{d}$ values with the corresponding lowest emission angles are listed here; see Fig.~\ref{fig:geometry} and Ref.~\cite{supplement} for their full 3D variation.\label{table:fitting}}
\begin{tabular}{l c c c c c c c}
\hline
\hline
\multicolumn{1}{c}{System} & $E_\mathrm{e}$\,[keV] & $E_\mathrm{d}$\,[eV]& \multicolumn{2}{c}{$\sigma_\mathrm{z}$}\,[barn] & \multicolumn{2}{c}{$\sigma_\mathrm{3D}$}\,[barn]& Ref.\\
\cline{4-7}
 & &($\gamma$,~$\delta$)~[$^\circ$] &$\xi$ & WMSE & $\xi$ & WMSE & \\
\hline
Graphene & 85--100 & 22.55~($0,~0$) &0.932 & 0.004& 0.938 &0.004 &\cite{susi_isotope_2016} \\
N--C$_2$   & 55--60  & 13.00~($55,~90$) & 0.276 & 0.459 & 0.394 &  0.103 & this work \\
\hline
\hline
\end{tabular}
\end{table*}

In Table~\ref{table:fitting} we compare the $z$-velocity dependent and 3D cross-section 
models, including the scaling factor and the residual error of the simultaneous fit to experimental data points at different electron energies~\cite{susi_isotope_2016}. While the full model makes only a small difference for pristine graphene, where the lowest threshold energy is reached at normal incidence to the plane, for the N--C$_2$ site not only is the required re-scaling smaller (from 0.27 to 0.39, a relative improvement of 44\%), the overall fitting error is reduced (from 0.42 to 0.02, a 21-fold decrease). In Figure~\ref{fig:pyrjumping}g, we can see that the 3D model also describes the trend in the data much better. However, even our complete elastic model can not explain the remarkably high experimentally observed jump rates: without the scaling factor $\xi$, the predicted rate would be negligible at these electron energies, while the N atoms are observed to jump many times per minute.

Since vibrations in all directions and the variation of $E_d$ as a function of emission angle and temperature are now accounted for in our complete theory, the remaining discrepancies between simulated and experimentally derived values have only one obvious source: the inaccuracy of DFT/MD in describing the energy required to eject the N atom from the N--C$_2$ site within the ground-state Born-Oppenheimer approximation.

Inelastic effects have been shown to be vital for explaining damage in non-metallic 2D materials~\cite{kotakoski_electron_2010,zan_control_2013,algara-siller_pristine_2013,susi_quantifying_2019, kretschmer_formation_2020}, but it is surprising they should also play such a role for point defects in metallic graphene. Recent theoretical work has proposed potential mechanisms and avenues for quantitative modeling~\cite{lingerfelt_nonadiabatic_2021}, but further advances, building upon the 3D foundation established here, will be needed before a full quantitative picture can be drawn.

Notably, the slightly incorrect slope of the theoretical cross section curve in Fig.~\ref{fig:pyrjumping}g with respect to the experimental data points could suggest an difference in the angular variation of the excited-state threshold energy. Alternatively, anharmonic effects not included in the phonon calculation of the velocities, or the variation of the threshold energy due to thermal fluctuations, might both also contribute to the somewhat different cross section as a function of electron energy. However, our results demonstrate that elastic energy transfer alone cannot be responsible for the observed beam-induced dynamics of pyridinic nitrogen impurity sites in graphene.

\begin{acknowledgments}
A.C., A.M., and T.S. were supported by the European Research Council (ERC) under the European Union’s Horizon 2020 research and innovation programme (Grant agreement No.~756277-ATMEN), and M.T. by the Austrian Science Fund (FWF) via project P 28322-N36. N.A.P. and M.J.V. gratefully acknowledge funding from the Belgian Fonds National de la Recherche Scientifique (FNRS) under grants PDR T.1077.15-1/7 and  T.0103.19-ALPS. M.J.V. acknowledges a FNRS sabbatical ``OUT'' grant at ICN2 Barcelona, as well as ULi{\`e}ge and the F\'ed\'eration Wallonie-Bruxelles (ARC DREAMS G.A. 21/25-01).
We gratefully acknowledge computational resources provided by the Vienna Scientific Cluster (VSC) and by the Consortium des Equipements de Calcul Intensif (CECI), funded by FRS-FNRS G.A. 2.5020.11; the Zenobe Tier-1 supercomputer funded by the Gouvernement Wallon G.A. 1117545; and by a PRACE-3IP DECI grants 2DSpin and Pylight on Beskow (G.A. 653838 of H2020). 
\end{acknowledgments}

\clearpage
\begin{widetext}
\setcounter{page}{1}
\setcounter{secnumdepth}{2}
\renewcommand{\theequation}{\Alph{section}.\arabic{equation}}
\setcounter{equation}{0}
\setcounter{figure}{0}
\renewcommand{\figurename}{Supplemental Figure}
\renewcommand{\thesection}{\Alph{section}}

\section{\label{lst:momtransfer}Derivation of 3D energy transfer}
Taking into account that the energy and momentum are conserved we have
\begin{eqnarray}
E_\mathrm{e}+E_\mathrm{n}&=&\tilde{E_{\mathrm{e}}}+\tilde{E_{\mathrm{n}}}\label{eq:enconserv}\;\; \mathrm{and}\\
\mathbf{p}_\mathrm{e}+\mathbf{p}_\mathrm{n}&=&\mathbf{\tilde{p}}_\mathrm{e}+\mathbf{\tilde{p}}_\mathrm{n},\label{eq:momconservation}
\end{eqnarray}
where the quantities $\tilde{E_{\mathrm{e}}}$, $\tilde{E_{\mathrm{n}}}$ and $\mathbf{\tilde{p}}_\mathrm{e}$, $\mathbf{\tilde{p}}_\mathrm{n}$ represent the total energies and momenta after the collision of the electron and nucleus, respectively. We can use the relativistic energy-momentum relationship to express $|\mathbf{\tilde{p}}_\mathrm{e}|$ as
\begin{equation}\label{eq:momafter}
|\mathbf{\tilde{p}}_\mathrm{e}|=\sqrt{\tilde{E_\mathrm{e}}\left(\tilde{E_\mathrm{e}}+2m c^2\right)/c^2}=\sqrt{\left(E_\mathrm{e}+E_\mathrm{n}-\tilde{E}_\mathrm{n}\right)\left(\left(E_\mathrm{e}+E_\mathrm{n}-\tilde{E}_\mathrm{n}\right)+2m c^2\right)/c^2},
\end{equation}
where $c$ is the speed of light and $m$ the rest mass of the electron. At this point, to reach an analytic solution we make the approximation $E_\mathrm{e}+E_\mathrm{n}-\tilde{E_\mathrm{n}}\approx E_\mathrm{e}$, as discussed in Ref.~\citenum{chirita_mihaila_influence_2019} and therefore $|\mathbf{p}_\mathrm{e}|=|\mathbf{\tilde{p}}_\mathrm{e}| $. After decomposing the momentum vectors in terms of their magnitude and direction (e.g. $\mathbf{p}_\mathrm{e}=|\mathbf{p}_\mathrm{e}|\mathbf{\hat{p}}_\mathrm{e}$) and using energy conservation we can rewrite equation~\ref{eq:momconservation} as follows:
\begin{equation}\label{eq:vecform}
\sqrt{E_\mathrm{e}\left(E_\mathrm{e}+2mc^2\right)/c^2}~\mathbf{\hat{p}}_\mathrm{e}+\sqrt{2ME_\mathrm{n}}~\mathbf{\hat{p}}_\mathrm{n}=\sqrt{E_\mathrm{e}\left(E_\mathrm{e}+2mc^2\right)/c^2}~\mathbf{\hat{\tilde{p}}}_\mathrm{e}+ \sqrt{2M\tilde{E}_\mathrm{n}}~\mathbf{\hat{\tilde{p}}}_\mathrm{n},
\end{equation}
where $\mathbf{\hat{p}}_\mathrm{e,n}$ and $\mathbf{\hat{\tilde{p}}}_\mathrm{e,n}$ give the direction of the electron and atom before and after the collision, respectively. We denote $|\mathbf{p}_\mathrm{e}|=p_\mathrm{e}=\sqrt{E_\mathrm{e}\left(E_\mathrm{e}+2m c^2\right)/c^2}$ and $|\mathbf{p}_\mathrm{n}|=p_\mathrm{n}=\sqrt{2ME_\mathrm{n}}=M v$ for brevity, where $M$ is the mass of the nucleus. Our goal is to determine the energy of the nucleus after the collision, $\tilde{E}_\mathrm{n}$. Shifting the third term to the left side of Eq.~\ref{eq:vecform} and then squaring results in
\begin{equation}\label{eq:emaxgeneral}
    \tilde{E}_\mathrm{n}= \frac{\left(p_\mathrm{e}\left(\mathbf{\hat{p}}_\mathrm{e}-\mathbf{\hat{\tilde{p}}}_\mathrm{e}\right)+p_\mathrm{n}\mathbf{\hat{p}}_\mathrm{n}\right)^2}{2M} =
    \frac{p_\mathrm{e}^2\left(\mathbf{\hat{p}}^2_\mathrm{e}+\mathbf{\hat{\tilde{p}}}^2_\mathrm{e}-2\mathbf{\hat{p}}_\mathrm{e}\cdot\mathbf{\hat{\tilde{p}}}_\mathrm{e}\right)+2p_\mathrm{e}p_\mathrm{n}\left(\mathbf{\hat{p}}_\mathrm{e}\cdot \mathbf{\hat{p}}_\mathrm{n}-\mathbf{\hat{\tilde{p}}}_\mathrm{e}\cdot \mathbf{\hat{p}}_\mathrm{n}\right)+p_\mathrm{n}^2\mathbf{\hat{p}}^2_\mathrm{n}}{2M}.
\end{equation}
This expression is the general representation of the elastic scattering of an electron from a moving target atom for arbitrary directions. We can recover the static limit~\cite{banhart_irradiation_1999} where the atom is considered to be at rest at the moment of impact by setting $p_\mathrm{n}=0$, $\mathbf{p}_\mathrm{e}=p_\mathrm{e}\left(0,0,1\right)$ in equation~\ref{eq:emaxgeneral}:
\begin{equation}\label{eq:emax2}
    \tilde{E}_\mathrm{n}= \frac{p_\mathrm{e}^2\left(\mathbf{\hat{p}}^2_\mathrm{e}+\mathbf{\hat{\tilde{p}}}^2_\mathrm{e}-2\mathbf{\hat{p}}_\mathrm{e}\cdot\mathbf{\hat{\tilde{p}}}_\mathrm{e}\right)}{2M}=\frac{p_\mathrm{e}^2\left(2-2\cos{\theta}\right)}{2M} \stackrel{\theta=\pi}{=} \frac{2E_\mathrm{e}\left(E_\mathrm{e}+2m c^2\right)}{M c^2},
\end{equation}
where the angle $\theta$ arising from the product between the initial and final electron momenta coincides with the spherical coordinate polar angle, and the last substitution yields the maximum energy transfer for electron backscattering.

If the atom is vibrating only in the out-of-plane $z$ direction at the moment of impact, we set $\mathbf{p}_\mathrm{e}=p_\mathrm{e}\left(0,0,1\right)$ for the incoming electron, and $\mathbf{p}_\mathrm{n}=p_\mathrm{n}\left(0,0,1\right)$ for the nucleus.
The product $\mathbf{\hat{p}}_\mathrm{n}\cdot \mathbf{\hat{p}}_\mathrm{e}$ yields unity since $\mathbf{\hat{p}}_\mathrm{n} = \mathbf{\hat{p}}_\mathrm{e} = (0,0,1)$, leading to $\mathbf{\hat{\tilde{p}}}_\mathrm{e}\cdot \mathbf{\hat{p}}_\mathrm{n}=\mathbf{\hat{\tilde{p}}}_\mathrm{e}\cdot \mathbf{\hat{p}}_\mathrm{e}=\cos{\theta}$. Substituting these terms in equation~\ref{eq:emaxgeneral} will result in~\cite{susi_quantifying_2019,chirita_mihaila_influence_2019}
\begin{eqnarray}\label{eq:emaxvz}
 \tilde{E}_\mathrm{n}\left(E_\mathrm{e},v_\mathrm{z},\theta\right)&=&\frac{p_\mathrm{e}^2\left(2-2\cos{\theta}\right)+2p_\mathrm{e}p_\mathrm{n}\left(1-\cos{\theta}\right)+p_\mathrm{n}^2}{2M} \notag \\
 &=&
 \frac{2\left(1-\cos{\theta}\right)\left(E_\mathrm{e}\left(E_\mathrm{e}+2m c^2\right)+Mv_\mathrm{z}c\sqrt{E_\mathrm{e}\left(E_\mathrm{e}+2m c^2\right)}\right)+\left(Mv_\mathrm{z}c\right)^2}{2Mc^2} \notag \\
 &\stackrel{\theta=\pi}{=}&
 \frac{\left(2\sqrt{E_\mathrm{e}\left(E_\mathrm{e}+2m c^2\right)}+Mv_\mathrm{z}c\right)^2}{2Mc^2}.
\end{eqnarray}

To derive a 3D description of the energy transfer we use a spherical coordinate system. Assuming that the incoming electron has an initial momentum along the $z$-axis, its direction vector is $\mathbf{p}_\mathrm{e}=p_\mathrm{e}\left(0,0,1\right)$. Since we know the initial momenta of the nucleus before the scattering occurs, we denote $\mathbf{p}_\mathrm{n}=M\left(v_\mathrm{x},v_\mathrm{y},v_\mathrm{z}\right)$. The electron can scatter at any angle after the collision $\mathbf{\tilde{p}}_\mathrm{e}=|\mathbf{p}_\mathrm{e}|\left(\sin{\theta}\cos{\varphi},\sin{\theta}\sin{\varphi},\cos{\theta}\right)$ and so can the atom  $\mathbf{\tilde{p}}_\mathrm{n}=|\mathbf{\tilde{p}}_\mathrm{n}|\left(\sin{\gamma}\cos{\delta},\sin{\gamma}\sin{\delta},\cos{\gamma}\right)$. Substituting these terms in Eq.~\ref{eq:emaxgeneral} will result in the 3D energy transfer:
\begin{eqnarray}
    \tilde{E}_\mathrm{n}\left(E_\mathrm{e},v_\mathrm{x},v_\mathrm{y},v_\mathrm{z},\theta, \varphi \right) &=& \frac{M^2\left(v_\mathrm{x}^2+v_\mathrm{y}^2+v_\mathrm{z}^2\right)}{2M}+ \frac{p_\mathrm{e}^2\left(2-2\cos{\theta}\right)+2p_\mathrm{e}M\left(v_\mathrm{z}-v_\mathrm{x}\sin{\theta}\cos{\varphi}-v_\mathrm{y}\sin{\theta}\sin{\varphi}-v_\mathrm{z}\cos{\theta}\right)}{2M}   \notag\\
    &=&
    \frac{M\left(v_\mathrm{x}^2+v_\mathrm{y}^2+v_\mathrm{z}^2\right)}{2}+\left(1-\cos{\theta}\right)\frac{E_{\mathrm{e}} \left(E_\mathrm{e}+2mc^2\right)+Mc v_\mathrm{z}\sqrt{E_\mathrm{e}\left(E_\mathrm{e}+2mc^2\right)}}{Mc^2}\notag\\
    &-&
    \frac{\sin{\theta}}{c}\left(v_\mathrm{x}\cos{\varphi}+v_\mathrm{y}\sin{\varphi}\right)\sqrt{E_\mathrm{e}\left(E_\mathrm{e}+2mc^2\right)}.
\end{eqnarray}{}\label{eq:emax3d}

While the scattering problem is naturally expressed in terms of the electron scattering angles $\theta$ and $\varphi$, the emission of the atom is here of greater interest. We thus need to derive a connection with the atom emission angles $\gamma$ and $\delta$ by decomposing the momentum conservation condition Eq.~\ref{eq:momconservation} to its Cartesian components as
\begin{eqnarray}
    p_\mathrm{n}^x &=& p_\mathrm{e}\sin{\theta}\cos{\varphi}+\tilde{p}_\mathrm{n}\sin{\gamma}\cos{\delta}, \label{eq:mom1}\\ 
    p_\mathrm{n}^y &=& p_\mathrm{e}\sin{\theta}\sin{\varphi}+\tilde{p}_\mathrm{n}\sin{\gamma}\sin{\delta}, \label{eq:mom2} \\ 
    p_\mathrm{n}^z + p_\mathrm{e} &=& p_\mathrm{e}\cos{\theta}+\tilde{p}_\mathrm{n}\cos{\gamma} \label{eq:mom3}.
\end{eqnarray}

We can solve Eqs.~\ref{eq:mom1}-\ref{eq:mom3} to derive expressions for the atom emission angles. To derive the angle $\delta$, we solve equation~\ref{eq:mom2} for $\tilde{p}_\mathrm{n}$
\begin{equation}
   \tilde{p}_\mathrm{n} = \frac{p_\mathrm{n}^y-p_\mathrm{e}\sin{\theta}\sin{\varphi}}{\sin{\gamma}\sin{\delta}},
\end{equation}
substitute this in equation~\ref{eq:mom1}
\begin{equation}
  p_\mathrm{n}^x - p_\mathrm{e}\sin{\theta}\cos{\varphi}= \frac{\left(p_\mathrm{n}^y-p_\mathrm{e}\sin{\theta}\sin{\varphi}\right)\cancel{\sin{\gamma}}\cos{\delta}}{\cancel{\sin{\gamma}}\sin{\delta}},
\end{equation}
and then solve for $\delta$:
\begin{eqnarray}\label{eq:delta}
  \delta\left( E_\mathrm{e},v_\mathrm{x},v_\mathrm{y},v_\mathrm{z},\varphi,\theta \right)
  &=&\arctan{\left(\frac{p_\mathrm{n}^y - p_\mathrm{e}\sin{\theta}\sin{\varphi}}{p_\mathrm{n}^x-p_\mathrm{e}\sin{\theta}\cos{\varphi}}\right)} \notag \\
  &=&\arctan{\left(\frac{Mv_\mathrm{y} - \sqrt{E_\mathrm{e}\left(E_\mathrm{e}+2mc^2\right)/c^2}\,\sin{\theta}\sin{\varphi}}{Mv_\mathrm{x}-\sqrt{E_\mathrm{e}\left(E_\mathrm{e}+2mc^2\right)/c^2}\,\sin{\theta}\cos{\varphi}}\right)} .
\end{eqnarray}
To obtain the atom emission angle $\gamma$, we solve equation~\ref{eq:mom1} for $\cos{\delta}$:
\begin{equation}\label{eq:a15}
    \cos{\delta}=\frac{p_\mathrm{n}^x - p_\mathrm{e}\sin{\theta}\cos{\varphi}}{\tilde{p}_\mathrm{n}\sin{\gamma}}.
\end{equation}
Using a trigonometric identity we can rewrite Eq.~\ref{eq:a15} as
\begin{equation}
\sin{\delta}=\sqrt{1-\frac{\left(p_\mathrm{n}^x - p_\mathrm{e}\sin{\theta}\cos{\varphi}\right)^2}{\tilde{p}^2_\mathrm{n}\sin{\gamma}^2}},
\end{equation}
and substitute the $\sin{\delta}$ term in equation~\ref{eq:mom2} to obtain
\begin{equation}
    p_\mathrm{n}^y -p_\mathrm{e}\sin{\theta}\sin{\varphi} = \sqrt{\tilde{p}^2_\mathrm{n}\sin{\gamma}^2-\left(p_\mathrm{n}^x - p_\mathrm{e}\sin{\theta}\cos{\varphi}\right)^2}.
\end{equation}
Next, we square both sides:
\begin{equation}\label{eq:A18}
   \tilde{p}^2_\mathrm{n}\sin{\gamma}^2 =\left(p_\mathrm{n}^x - p_\mathrm{e}\sin{\theta}\cos{\varphi}\right)^2+\left(p_\mathrm{n}^y -p_\mathrm{e}\sin{\theta}\sin{\varphi}\right)^2,
\end{equation}
solve equation~\ref{eq:mom3} for $\tilde{p}_\mathrm{n}$
\begin{equation}
   \tilde{p}_\mathrm{n}  =\frac{p_\mathrm{n}^z + p_\mathrm{e}\left( 1-\cos{\theta} \right)}{\cos{\gamma}},
\end{equation}
and substitute it in Eq.~\ref{eq:A18} to solve for $\gamma$
\begin{eqnarray}
    \gamma\left( E_\mathrm{e},v_\mathrm{x},v_\mathrm{y},v_\mathrm{z},\varphi,\theta \right) 
    &=& \arctan\left(\frac{\sqrt{\left(p_\mathrm{n}^x-p_\mathrm{e}\sin{\theta}\cos{\varphi} \right)^2+ \left(p_\mathrm{n}^y-p_\mathrm{e}\sin{\theta}\sin{\varphi} \right)^2 }}{p_\mathrm{n}^z+p_\mathrm{e}\left( 1-\cos{\theta} \right)}\right)\notag \\
    &=& \arctan\left(\left[\left(Mv_\mathrm{x}-\sqrt{E_\mathrm{e}\left(E_\mathrm{e}+2mc^2\right)/c^2}\,\sin{\theta}\cos{\varphi} \right)^2 \right.\right. \notag \\
    &+&\left. \left(Mv_\mathrm{y}-\sqrt{E_\mathrm{e}\left(E_\mathrm{e}+2mc^2\right)/c^2}\,\sin{\theta}\sin{\varphi}\right)^2\right]^{1/2} \notag \\
    &/&\left.\left(Mv_\mathrm{z}+\sqrt{E_\mathrm{e}\left(E_\mathrm{e}+2mc^2\right)/c^2}\,\left( 1-\cos{\theta} \right)\right)\right)\label{eq:gamma}.
\end{eqnarray}

\section{Atomic mean-square velocities}\label{lst:phononvelocities}
The density of states and vibrational properties are calculated by finding the interatomic force constants of a graphene-like super-cell containing 72 atoms. For the N--C$_2$ impurity, we replace a single C atom with N and remove a nearest-neighbor C to create the nitrogen-vacancy center. We use density functional perturbation theory as implemented in the \textsc{Abinit} software package~\cite{romero_abinit_2020}, with norm-conserving pseudo-potentials generated with the ONCVPSP code~\cite{hamann_optimized_2013, setten_pseudodojo_2018}, a plane-wave basis set, and the PBE exchange-correlation functional~\cite{perdew_generalized_1996}. After calculating the phonon dispersion, a displacement-weighted phonon density of states (pDOS) is constructed for each atom using a Gaussian smearing of 0.6 meV, and populated with Bose-Einstein thermal factors at 300~K. Within a harmonic model, the mean-square velocity is related directly to the mean-square displacement by the mode frequency~\cite{lee_ab_1995}. 

Using the interatomic force constants, we calculate the phonon band structure, atom-projected pDOS (orange and blue), and total pDOS (black lines) in Suppl. Fig.~\ref{sifig:dos_bands}. We note that our calculations of the N--C$_2$ impurity gave rise to a high-frequency state located both on the impurity and on the edge of the super-cell (halfway between image impurities in periodic boundary conditions). We believe that the finite size of our simulation cell gives rise to an artificial localization. With additional supercell convergence, this mode should merge with those of the remaining C atoms. 

At room temperature its impact is minimal: for our calculated mean-square velocities of the atoms for the N--C$_2$ site in graphene displayed in Suppl. Fig.~\ref{sifig:N-C2_velocities}, we artificially removed this high-frequency mode from our calculations of the mean-square displacement and velocity. However, we note that the magnitude of the velocities for the nearest-neighbor C atoms change by around 8.9\% due to the removal of this high-frequency mode and thus our results are marginally affected by the removal of this mode.

\begin{figure}[h]
    \centering
    \includegraphics[width=0.6\linewidth]{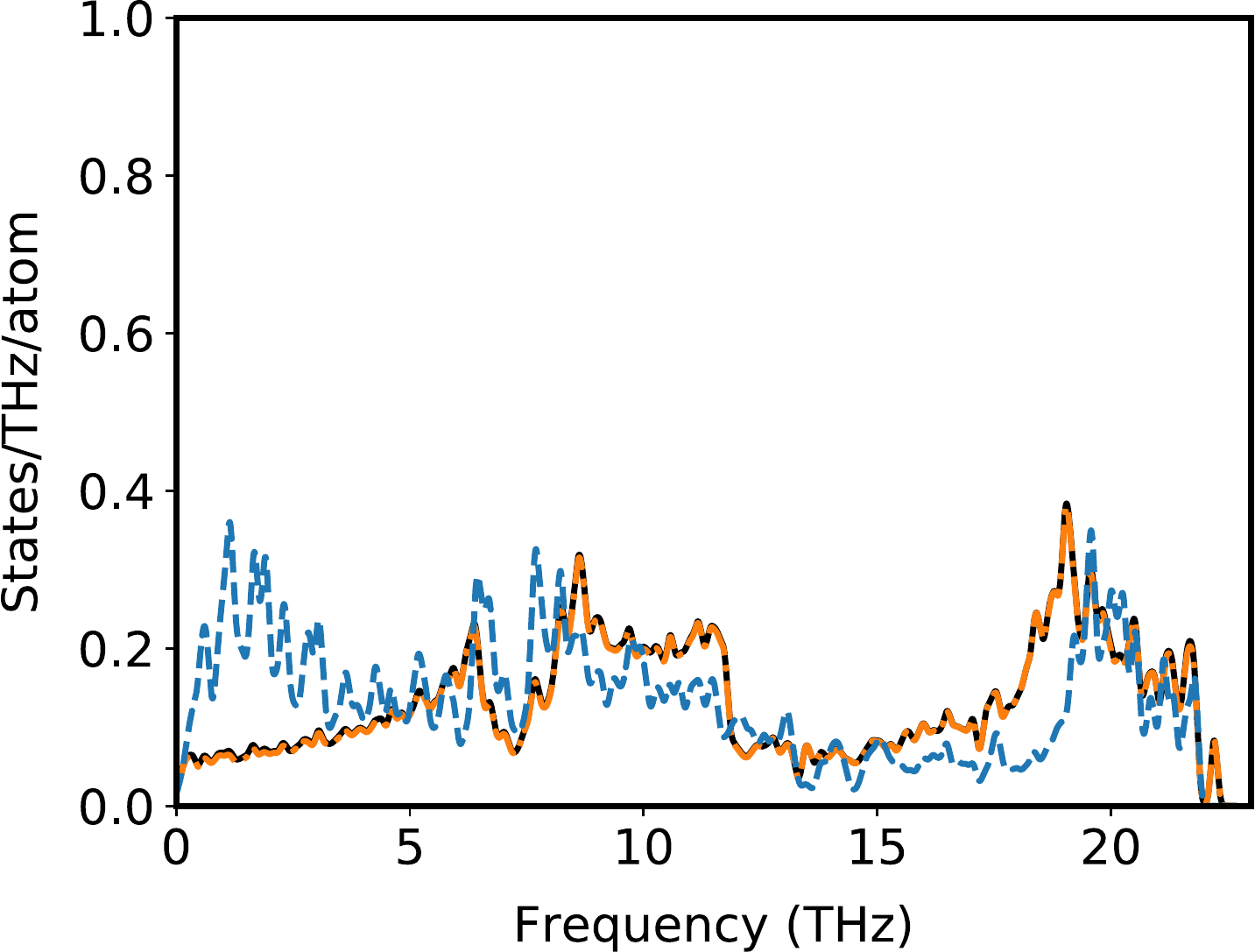}
    \caption{The atom projected and total phonon density of states (pDOS) of the N--C$_2$ site in graphene calculated from the interatomic force constants. For the plot, we use an orange line for C, a blue dashed line for the N atom, and a black line for the total pDOS. In both cases, we show the average pDOS contribution \emph{per atom}.}
    \label{sifig:dos_bands}
\end{figure}

\clearpage

\begin{figure}[ht]
    \centering
    \includegraphics[width=0.95\linewidth]{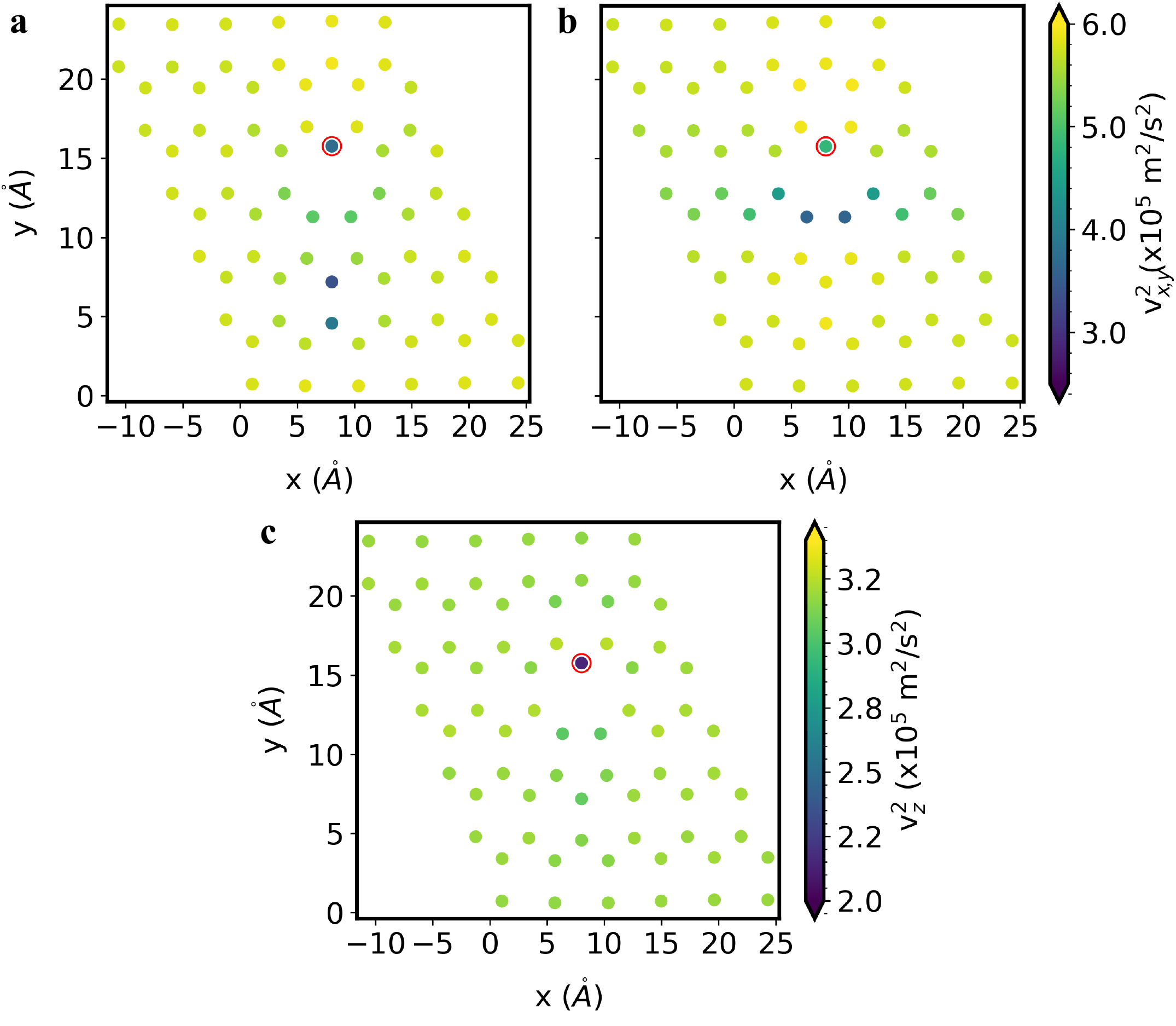}
    \caption{Atomic mean-square velocities of the N--C$_2$ site (N atom highlighted by the red circle) in graphene calculated from the phonon density of states. (a) $\overline{v_x^2}$, (b) $\overline{v_y^2}$, and (c) $\overline{v_x^2}$ color-coded in units of m$^2$/s$^2$.}
    \label{sifig:N-C2_velocities}
\end{figure}

\clearpage

\section{3D threshold energies for pyridinic nitrogen}\label{lst:3D_Td_Si_N}

The effective area for possible pyridinic N atom (N--C$_2$) jumps and the displacement threshold energy $E_\mathrm{d}$ is shown in Suppl. Fig.~\ref{sifig:N_Si_Td}. This area is the result of the numerical interpolation between the $T_\mathrm{d}$ values for that particular angular range. 

\begin{figure}[hb]
    \centering
    \includegraphics[width=0.7\linewidth]{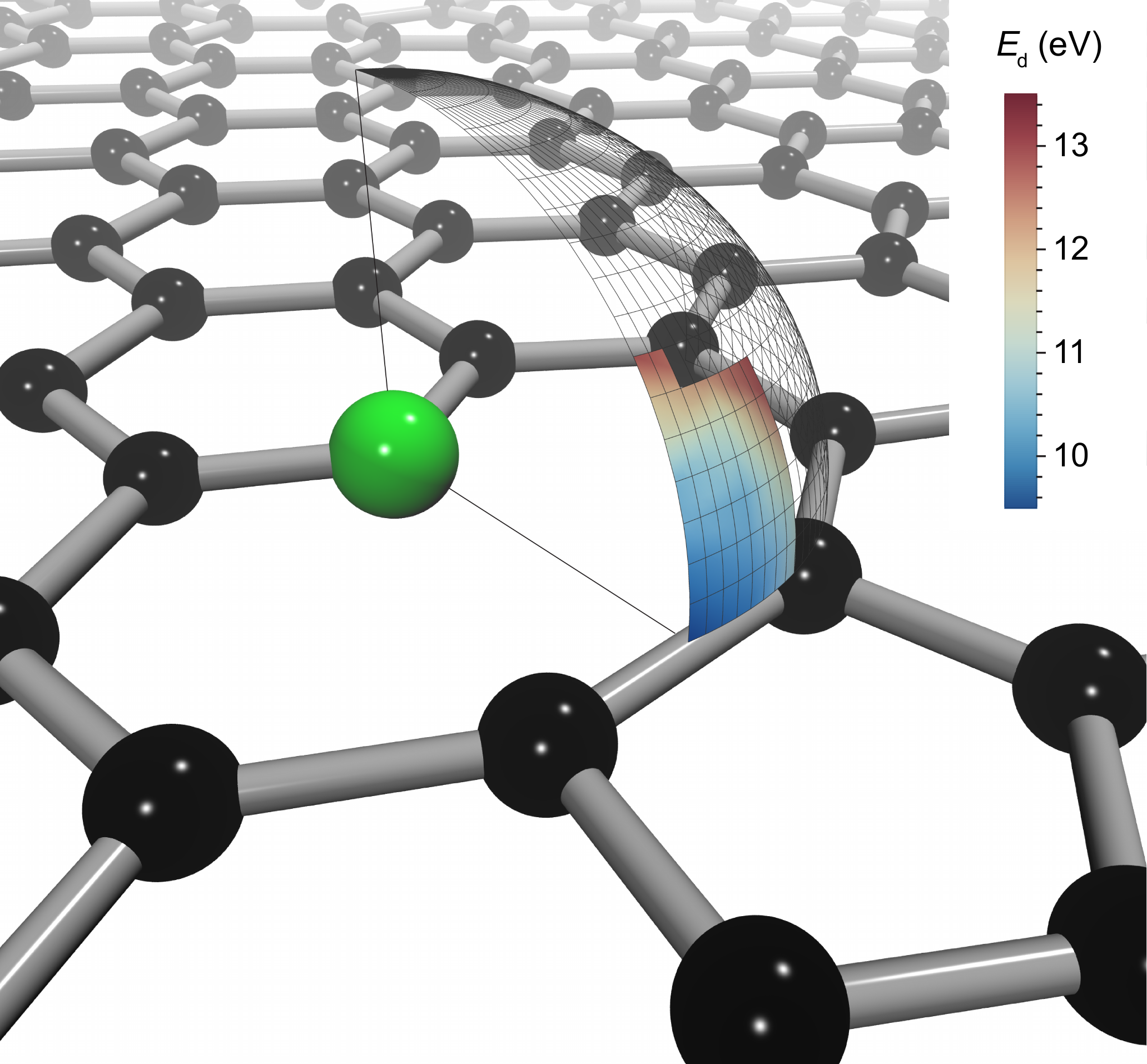}
    \caption{Variation of $E_\mathrm{d}$ for the jumping of the N--C$_2$ (N atom shown in green) in graphene estimated with DFT/MD and numerically interpolated. The direction across the vacancy is aligned with the positive $y$ axis.}
    \label{sifig:N_Si_Td}
\end{figure}

\clearpage

\section{Theoretical description of cross sections}\label{lst:cross-section}
McKinley and Feshbach~\cite{mckinley_coulomb_1948} derived a cross-section formula for mid-atomic number elements 

\begin{equation}\label{eq:mckinley}
\sigma =\sigma_\mathrm{R}\left[ 1-\beta^2\sin^2{\theta/2}+\pi Z \alpha \beta \sin{\theta/2}\left(1-\sin{\theta/2}\right) \right],
\end{equation}
where $\theta$ is the electron scattering angle, $Z$ is the atomic number, $\alpha=\frac{1}{4\pi \epsilon_0}\frac{e^2}{\hbar c}$ is the fine structure constant, $\beta=v/c=\sqrt{1-1/\left(U/mc^2+1\right)^2}$ is the ratio of the electron velocity with speed of light for acceleration voltage $U$, and $\sigma_\mathrm{R}$ is the classical Rutherford scattering cross section

\begin{equation}
\sigma_\mathrm{R}=\left(\frac{Ze^2}{4\pi\epsilon_02mc^2}\right)^2\frac{1-\beta^2}{\beta^4}\csc^4(\theta/2),    
\end{equation}
where $\epsilon_0$ is the vacuum permittivity.

In the static approximation, one would integrate Eq.~\ref{eq:mckinley} over $\theta$ and only count events where the transferred energy $\tilde{E}_\mathrm{n}\left(E_\mathrm{e},\theta \right)$ exceeds the displacement threshold energy of the atom $E_\mathrm{d}$~\cite{zobelli_electron_2007}
\begin{equation}
    \sigma_\mathrm{static}=\int\displaylimits_{\theta=0}^{\theta=\pi}\mathcal{H}(\tilde{E}_\mathrm{n}\left(E_\mathrm{e},\theta \right)-E_\mathrm{d})\,\sigma\left( E_\mathrm{e},\theta \right)\mathrm{d}\theta.
\end{equation}
The Heaviside function $\mathcal{H}$ will be taking the values 1 or 0, depending on whether the transferred energy $\tilde{E}_\mathrm{n}\left(E_\mathrm{e},\theta \right)$ exceeds the displacement threshold energy $E_\mathrm{d}$ or not, respectively. 

In the non-static approach discussed in~\cite{meyer_accurate_2012,susi_isotope_2016,chirita_mihaila_influence_2019}, the out-of-plane velocity $v_\mathrm{z}$ of the target atom comes in addition to the electron scattering angle $\theta$, and so the cross-section integral needs to be sampled over both $\theta$ and $v_\mathrm{z}$ where the transferred energy exceeds the displacement threshold of the target atom. The probability distribution of velocities of the target atom is given by a normal distribution:

\begin{equation}\label{eq:distribution}
   P(v_i,\overline{v^2_i})=\frac{1}{\sqrt{2\pi\overline{v^2_i}}}\exp\left( \frac{-v^2_i}{2\overline{v^2_i}}\right), 
\end{equation}
 where the mean squared velocity $\overline{v^2_i}$ is the variance. The temperature dependence of the displacement cross-section can also be quantified by the variation of the displacement threshold energy, which is mainly caused by the vibrational motion of the neighboring carbon atoms. If the neighbors happen to have momentum in the opposite direction of the ejection of the C atom, the force components will point in the direction of the ejection and so the displacement threshold will be lower. Should they vibrate in the direction of the ejecting C atom, the displacement threshold energy increases, as was discussed in a previous work by Chirita et. al~\cite{chirita_mihaila_influence_2019}.

To quantify the emission of atoms in certain directions and predict particular events, such as the pyridinic N jump across the vacancy, one has to be able to limit the cross-section to an angular sector of interest. 
This can be generalized to all three dimensions to predict the displacement cross-sections in any arbitrary direction, taking into consideration atom velocities prior to the scattering event and changes in $E_\mathrm{d}$ caused by thermal perturbations or by different atomic emission angles. This allows us to integrate over Eq.~\ref{eq:mckinley} over an arbitrary set of angles and velocities, where the transferred energy $\tilde{E}_\mathrm{n}$ exceeds the displacement threshold $\xi E_\mathrm{d}$:

\begin{equation}\label{eq:sigma3dApp}
\sigma_\mathrm{3D}= \int\int\prod_{i=\{x,y,z\}}\int\displaylimits_{-v_i^\mathrm{max}}^{v_i^\mathrm{max}}
P(v_i,\overline{v_i^2})\,\mathcal{H}(\tilde{E}_\mathrm{n}-\xi E_\mathrm{d}\left(\gamma,\delta\right))\,\sigma(E_\mathrm{e},\theta)\,\mathrm{d}v_i\,\sin{\theta}\,\mathrm{d}\theta\,\mathrm{d}\varphi,
\end{equation}
where $P(v_i,\overline{v^2_i})$ with $i={x,y,z}$ are the normal distributions of velocities of the target atom with mean-square velocities (variance) $\overline{v_i^2}$ derived from the phonon density of states~\cite{susi_isotope_2016} and integrated over $\pm v_i^{max}$ covering the variation of velocities, 
$\mathcal{H}$ is the Heaviside step function ensuring the transferred energy $\tilde{E}_\mathrm{n}$ exceeds $E_\mathrm{d}$ and $\xi$ is a fitting factor multiplying the simulated $E_\mathrm{d}$ to help match the experimentally observed displacement cross-sections $\sigma_\mathrm{exp}$. As a measure of the goodness of the fit and for fitting $\xi$ we calculate a weighted mean-square error (WMSE).

\section{Numerical cross section integrals}\label{lst:integration}
The demanding five-dimensional cross-section integral of Eq.~\ref{eq:sigma3dApp} must be integrated using numerical methods, and we did not find ``brute-force" approaches to yield a converged result in any reasonable time. We thus trialed several different adaptive algorithms and found Monte Carlo methods to be numerically imprecise. We thus opted for the \texttt{GlobalAdaptive} algorithm~\cite{krommer_computational_1998} as implemented in Mathematica~\cite{Mathematica}. To ensure the numerical precision of the integrals to two decimal places we used the option \texttt{PrecisionGoal=2}, and for a finer subdivision of the integration region, especially in areas where the cross-section is small, we opted for \texttt{MinRecursion=4}. To simulate a cross-section for a single voltage point to this degree of accuracy usually takes one minute on a modern CPU, and one can parallelize the computation over available cores using \texttt{ParallelTable}.

To describe the dependence of $E_d$ on the atom emission angles we used the interpolation function $E_\mathrm{d}\left(\gamma,\delta\right)$ defined piece-wise as follows:
\begin{equation}
    {E_\mathrm{d}}\left(\gamma,\delta\right)=
                                        \begin{cases} 
                                      E_\mathrm{d}\left(\gamma,\delta\right) &\gamma_\mathrm{1}\leq \gamma \leq \gamma_\mathrm{2}$ and $\delta_\mathrm{1} \leq \delta \leq \delta_\mathrm{2} \\
                                        \infty 
                                        \end{cases}.
                                        \label{eq:interpolation}
\end{equation}

For graphene, we simulated C atom displacements for emission angles $0^\circ\leq \gamma \leq 30^\circ$ and $0^\circ \leq \delta \leq 60^\circ$ due to the sixfold symmetry of the ejection geometry. Energy transfers for $\gamma > 30^\circ$ are negligible and do not contribute to the cross-section at modest electron energies as the displacement threshold energy cannot be reached (see Suppl. Fig.~\ref{sifig:energytransfers}). For the N-doped graphene, we ran N displacement simulation in the range of $250^\circ\leq \delta \leq 290^\circ$ and $55^\circ \leq \gamma \leq 90^\circ$ (two-fold symmetry). However, the cross-section is determined (up to 98\%) by angles $\gamma < 70^\circ$, highlighting how little energy can be transferred to angles closer to the plane.

\begin{figure}[h]
    \centering
    \includegraphics[width=0.95\linewidth]{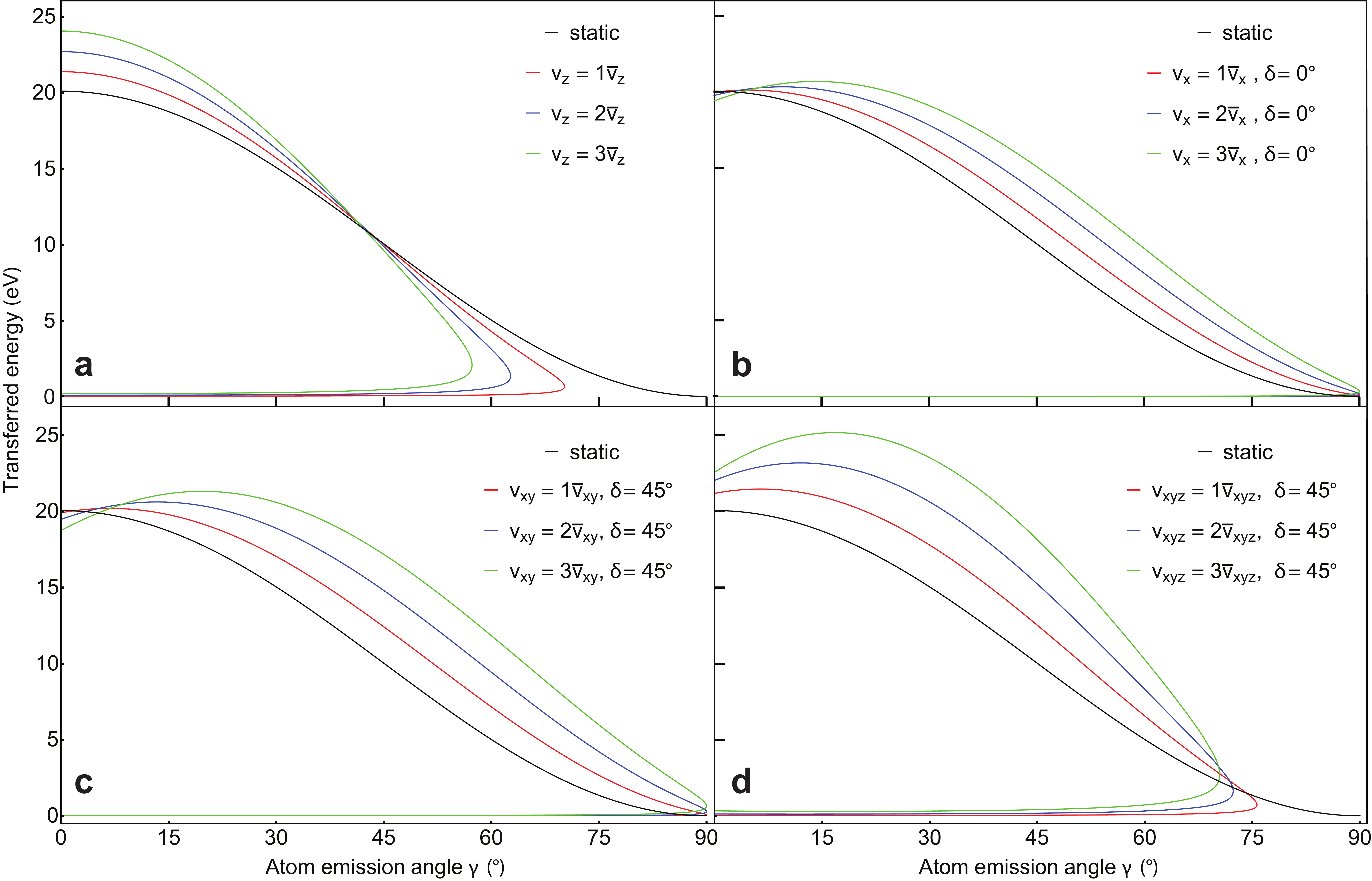}
    \caption{Transferred energy $\tilde{E}_\mathrm{n}$ as a function of atom polar emission angle $\gamma$ for $^{12}$C graphene for different initial velocities and azimuthal emission angle $\delta$. The 3D model is compared to the static approximation, where no initial velocity of the atom is taken into consideration. For the non-static cases, the energy curves are plotted for 1--3 standard deviations from the velocity mean in each of the (a) $z$ ($\overline{v_z}$), (b) $x$ ($\overline{v_x}$), (c) $xy$ ($\overline{v_x}$ and $\overline{v_y}$), and (d) $xyz$ ($\overline{v_x}$, $\overline{v_y}$, $\overline{v_z}$) directions, plotted for (b) $\delta=0^\circ$ or (c,d) $\delta=45^\circ$ against the polar atom emission angle $\gamma$.}
    \label{sifig:energytransfers}
\end{figure}

\section{Analysis of experimental cross sections}\label{lst:analysis}
We quantify the electron-beam-induced knock-on effects by estimating the displacement cross-section. We record a series of images over a specific scan area with a known electron dose. Whenever a jump or displacement event happens, we calculate the accumulated dose. Since this is a stochastic process, the event doses will be Poisson-distributed. Therefore the expectation value of the corresponding exponential distribution can be fitted as in Fig.~\ref{fig:pyrjumping} and the experimental cross-section evaluated as

\begin{equation}\label{eq:sigmaexp}
    \sigma_\mathrm{exp}=\frac{1}{\rho \lambda},
\end{equation}
where $\rho$ is the atomic surface density of the material in units of atoms/$\mathrm{m}^2$ and $\lambda$ is the Poisson expectation value.

\end{widetext}


\begin{thebibliography}{38}%
\makeatletter
\providecommand \@ifxundefined [1]{%
 \@ifx{#1\undefined}
}%
\providecommand \@ifnum [1]{%
 \ifnum #1\expandafter \@firstoftwo
 \else \expandafter \@secondoftwo
 \fi
}%
\providecommand \@ifx [1]{%
 \ifx #1\expandafter \@firstoftwo
 \else \expandafter \@secondoftwo
 \fi
}%
\providecommand \natexlab [1]{#1}%
\providecommand \enquote  [1]{``#1''}%
\providecommand \bibnamefont  [1]{#1}%
\providecommand \bibfnamefont [1]{#1}%
\providecommand \citenamefont [1]{#1}%
\providecommand \href@noop [0]{\@secondoftwo}%
\providecommand \href [0]{\begingroup \@sanitize@url \@href}%
\providecommand \@href[1]{\@@startlink{#1}\@@href}%
\providecommand \@@href[1]{\endgroup#1\@@endlink}%
\providecommand \@sanitize@url [0]{\catcode `\\12\catcode `\$12\catcode
  `\&12\catcode `\#12\catcode `\^12\catcode `\_12\catcode `\%12\relax}%
\providecommand \@@startlink[1]{}%
\providecommand \@@endlink[0]{}%
\providecommand \url  [0]{\begingroup\@sanitize@url \@url }%
\providecommand \@url [1]{\endgroup\@href {#1}{\urlprefix }}%
\providecommand \urlprefix  [0]{URL }%
\providecommand \Eprint [0]{\href }%
\providecommand \doibase [0]{http://dx.doi.org/}%
\providecommand \selectlanguage [0]{\@gobble}%
\providecommand \bibinfo  [0]{\@secondoftwo}%
\providecommand \bibfield  [0]{\@secondoftwo}%
\providecommand \translation [1]{[#1]}%
\providecommand \BibitemOpen [0]{}%
\providecommand \bibitemStop [0]{}%
\providecommand \bibitemNoStop [0]{.\EOS\space}%
\providecommand \EOS [0]{\spacefactor3000\relax}%
\providecommand \BibitemShut  [1]{\csname bibitem#1\endcsname}%
\let\auto@bib@innerbib\@empty
\bibitem [{\citenamefont {Hawkes}(2009)}]{hawkes_aberration_2009}%
  \BibitemOpen
  \bibfield  {author} {\bibinfo {author} {P.~W. Hawkes},\ }\href {\doibase
  10.1098/rsta.2009.0004} {\bibfield  {journal} {\bibinfo  {journal}
  {Philosophical Transactions of the Royal Society of London A: Mathematical,
  Physical and Engineering Sciences}\ }\textbf {\bibinfo {volume} {367}},\
  \bibinfo {pages} {3637} (\bibinfo {year} {2009})}\BibitemShut {NoStop}%
\bibitem [{\citenamefont {Egerton}(2019)}]{egerton_radiation_2019}%
  \BibitemOpen
  \bibfield  {author} {\bibinfo {author} {R.~F. Egerton},\ }\href {\doibase
  https://doi.org/10.1016/j.micron.2019.01.005} {\bibfield  {journal} {\bibinfo
   {journal} {Micron}\ }\textbf {\bibinfo {volume} {119}},\ \bibinfo {pages}
  {72} (\bibinfo {year} {2019})}\BibitemShut {NoStop}%
\bibitem [{\citenamefont {Banhart}(1999)}]{banhart_irradiation_1999}%
  \BibitemOpen
  \bibfield  {author} {\bibinfo {author} {F.~Banhart},\ }\href {\doibase
  10.1088/0034-4885/62/8/201} {\bibfield  {journal} {\bibinfo  {journal} {Rep.
  Prog. Phys.}\ }\textbf {\bibinfo {volume} {62}},\ \bibinfo {pages} {1181}
  (\bibinfo {year} {1999})}\BibitemShut {NoStop}%
\bibitem [{\citenamefont {Meyer}\ \emph {\emph{and others}}(2008)\citenamefont
  {Meyer}, \citenamefont {Kisielowski}, \citenamefont {Erni}, \citenamefont
  {Rossell}, \citenamefont {Crommie},\ and\ \citenamefont
  {Zettl}}]{meyer_direct_2008}%
  \BibitemOpen
  \bibfield  {author} {\bibinfo {author} {J.~C. Meyer}, \bibinfo {author}
  {C.~Kisielowski}, \bibinfo {author} {R.~Erni}, \bibinfo {author} {M.~D.
  Rossell}, \bibinfo {author} {M.~F. Crommie},\ and\ \bibinfo {author}
  {A.~Zettl},\ }\href {\doibase 10.1021/nl801386m} {\bibfield  {journal}
  {\bibinfo  {journal} {Nano Letters}\ }\textbf {\bibinfo {volume} {8}},\
  \bibinfo {pages} {3582} (\bibinfo {year} {2008})}\BibitemShut {NoStop}%
\bibitem [{\citenamefont {Kotakoski}\ \emph {\emph{and
  others}}(2011)\citenamefont {Kotakoski}, \citenamefont {Meyer}, \citenamefont
  {Kurasch}, \citenamefont {Santos-Cottin}, \citenamefont {Kaiser},\ and\
  \citenamefont {Krasheninnikov}}]{kotakoski_stone-wales-type_2011}%
  \BibitemOpen
  \bibfield  {author} {\bibinfo {author} {J.~Kotakoski}, \bibinfo {author}
  {J.~C. Meyer}, \bibinfo {author} {S.~Kurasch}, \bibinfo {author}
  {D.~Santos-Cottin}, \bibinfo {author} {U.~Kaiser},\ and\ \bibinfo {author}
  {A.~V. Krasheninnikov},\ }\href {\doibase 10.1103/PhysRevB.83.245420}
  {\bibfield  {journal} {\bibinfo  {journal} {Phys. Rev. B}\ }\textbf {\bibinfo
  {volume} {83}},\ \bibinfo {pages} {245420} (\bibinfo {year}
  {2011})}\BibitemShut {NoStop}%
\bibitem [{\citenamefont {Susi}\ \emph {\emph{and others}}(2014)\citenamefont
  {Susi}, \citenamefont {Kotakoski}, \citenamefont {Kepaptsoglou},
  \citenamefont {Mangler}, \citenamefont {Lovejoy}, \citenamefont {Krivanek},
  \citenamefont {Zan}, \citenamefont {Bangert}, \citenamefont {Ayala},
  \citenamefont {Meyer},\ and\ \citenamefont
  {Ramasse}}]{susi_siliconcarbon_2014}%
  \BibitemOpen
  \bibfield  {author} {\bibinfo {author} {T.~Susi}, \bibinfo {author}
  {J.~Kotakoski}, \bibinfo {author} {D.~Kepaptsoglou}, \bibinfo {author}
  {C.~Mangler}, \bibinfo {author} {T.~C. Lovejoy}, \bibinfo {author} {O.~L.
  Krivanek}, \bibinfo {author} {R.~Zan}, \bibinfo {author} {U.~Bangert},
  \bibinfo {author} {P.~Ayala}, \bibinfo {author} {J.~C. Meyer},\ and\ \bibinfo
  {author} {Q.~Ramasse},\ }\href {\doibase 10.1103/PhysRevLett.113.115501}
  {\bibfield  {journal} {\bibinfo  {journal} {Phys. Rev. Lett.}\ }\textbf
  {\bibinfo {volume} {113}},\ \bibinfo {pages} {115501} (\bibinfo {year}
  {2014})}\BibitemShut {NoStop}%
\bibitem [{\citenamefont {Susi}\ \emph {\emph{and
  others}}(2017{\natexlab{a}})\citenamefont {Susi}, \citenamefont {Meyer},\
  and\ \citenamefont {Kotakoski}}]{susi_manipulating_2017}%
  \BibitemOpen
  \bibfield  {author} {\bibinfo {author} {T.~Susi}, \bibinfo {author}
  {J.~Meyer},\ and\ \bibinfo {author} {J.~Kotakoski},\ }\href {\doibase
  10.1016/j.ultramic.2017.03.005} {\bibfield  {journal} {\bibinfo  {journal}
  {Ultramicroscopy}\ }\textbf {\bibinfo {volume} {180}},\ \bibinfo {pages}
  {163} (\bibinfo {year} {2017}{\natexlab{a}})}\BibitemShut {NoStop}%
\bibitem [{\citenamefont {Dyck}\ \emph {\emph{and others}}(2017)\citenamefont
  {Dyck}, \citenamefont {Kim}, \citenamefont {Kalinin},\ and\ \citenamefont
  {Jesse}}]{dyck_placing_2017}%
  \BibitemOpen
  \bibfield  {author} {\bibinfo {author} {O.~Dyck}, \bibinfo {author} {S.~Kim},
  \bibinfo {author} {S.~V. Kalinin},\ and\ \bibinfo {author} {S.~Jesse},\
  }\href {\doibase 10.1063/1.4998599} {\bibfield  {journal} {\bibinfo
  {journal} {Applied Physics Letters}\ }\textbf {\bibinfo {volume} {111}},\
  \bibinfo {pages} {113104} (\bibinfo {year} {2017})}\BibitemShut {NoStop}%
\bibitem [{\citenamefont {Tripathi}\ \emph {\emph{and
  others}}(2018)\citenamefont {Tripathi}, \citenamefont {Mittelberger},
  \citenamefont {Pike}, \citenamefont {Mangler}, \citenamefont {Meyer},
  \citenamefont {Verstraete}, \citenamefont {Kotakoski},\ and\ \citenamefont
  {Susi}}]{tripathi_electron-beam_2018}%
  \BibitemOpen
  \bibfield  {author} {\bibinfo {author} {M.~Tripathi}, \bibinfo {author}
  {A.~Mittelberger}, \bibinfo {author} {N.~A. Pike}, \bibinfo {author}
  {C.~Mangler}, \bibinfo {author} {J.~C. Meyer}, \bibinfo {author} {M.~J.
  Verstraete}, \bibinfo {author} {J.~Kotakoski},\ and\ \bibinfo {author}
  {T.~Susi},\ }\href {\doibase 10.1021/acs.nanolett.8b02406} {\bibfield
  {journal} {\bibinfo  {journal} {Nano Letters}\ }\textbf {\bibinfo {volume}
  {18}},\ \bibinfo {pages} {5319} (\bibinfo {year} {2018})}\BibitemShut
  {NoStop}%
\bibitem [{\citenamefont {Susi}\ \emph {\emph{and others}}(2019)\citenamefont
  {Susi}, \citenamefont {Meyer},\ and\ \citenamefont
  {Kotakoski}}]{susi_quantifying_2019}%
  \BibitemOpen
  \bibfield  {author} {\bibinfo {author} {T.~Susi}, \bibinfo {author} {J.~C.
  Meyer},\ and\ \bibinfo {author} {J.~Kotakoski},\ }\href {\doibase
  10.1038/s42254-019-0058-y} {\bibfield  {journal} {\bibinfo  {journal} {Nature
  Reviews Physics}\ }\textbf {\bibinfo {volume} {1}},\ \bibinfo {pages} {397}
  (\bibinfo {year} {2019})}\BibitemShut {NoStop}%
\bibitem [{\citenamefont {Meyer}\ \emph {\emph{and others}}(2012)\citenamefont
  {Meyer}, \citenamefont {Eder}, \citenamefont {Kurasch}, \citenamefont
  {Skakalova}, \citenamefont {Kotakoski}, \citenamefont {Park}, \citenamefont
  {Roth}, \citenamefont {Chuvilin}, \citenamefont {Eyhusen}, \citenamefont
  {Benner}, \citenamefont {Krasheninnikov},\ and\ \citenamefont
  {Kaiser}}]{meyer_accurate_2012}%
  \BibitemOpen
  \bibfield  {author} {\bibinfo {author} {J.~C. Meyer}, \bibinfo {author}
  {F.~Eder}, \bibinfo {author} {S.~Kurasch}, \bibinfo {author} {V.~Skakalova},
  \bibinfo {author} {J.~Kotakoski}, \bibinfo {author} {H.~J. Park}, \bibinfo
  {author} {S.~Roth}, \bibinfo {author} {A.~Chuvilin}, \bibinfo {author}
  {S.~Eyhusen}, \bibinfo {author} {G.~Benner}, \bibinfo {author} {A.~V.
  Krasheninnikov},\ and\ \bibinfo {author} {U.~Kaiser},\ }\href {\doibase
  10.1103/PhysRevLett.108.196102} {\bibfield  {journal} {\bibinfo  {journal}
  {Phys. Rev. Lett.}\ }\textbf {\bibinfo {volume} {108}},\ \bibinfo {pages}
  {196102} (\bibinfo {year} {2012})}\BibitemShut {NoStop}%
\bibitem [{\citenamefont {Egerton}(2013)}]{egerton_beam-induced_2013}%
  \BibitemOpen
  \bibfield  {author} {\bibinfo {author} {R.~Egerton},\ }\href {\doibase
  10.1017/S1431927612014274} {\bibfield  {journal} {\bibinfo  {journal}
  {Microscopy and Microanalysis}\ }\textbf {\bibinfo {volume} {19}},\ \bibinfo
  {pages} {479} (\bibinfo {year} {2013})}\BibitemShut {NoStop}%
\bibitem [{\citenamefont {Susi}\ \emph {\emph{and others}}(2016)\citenamefont
  {Susi}, \citenamefont {Hofer}, \citenamefont {Argentero}, \citenamefont
  {Leuthner}, \citenamefont {Pennycook}, \citenamefont {Mangler}, \citenamefont
  {Meyer},\ and\ \citenamefont {Kotakoski}}]{susi_isotope_2016}%
  \BibitemOpen
  \bibfield  {author} {\bibinfo {author} {T.~Susi}, \bibinfo {author}
  {C.~Hofer}, \bibinfo {author} {G.~Argentero}, \bibinfo {author} {G.~T.
  Leuthner}, \bibinfo {author} {T.~J. Pennycook}, \bibinfo {author}
  {C.~Mangler}, \bibinfo {author} {J.~C. Meyer},\ and\ \bibinfo {author}
  {J.~Kotakoski},\ }\href {\doibase 10.1038/ncomms13040} {\bibfield  {journal}
  {\bibinfo  {journal} {Nature Communications}\ }\textbf {\bibinfo {volume}
  {7}},\ \bibinfo {pages} {13040} (\bibinfo {year} {2016})}\BibitemShut
  {NoStop}%
\bibitem [{\citenamefont {Chirita~Mihaila}\ \emph {\emph{and
  others}}(2019)\citenamefont {Chirita~Mihaila}, \citenamefont {Susi},\ and\
  \citenamefont {Kotakoski}}]{chirita_mihaila_influence_2019}%
  \BibitemOpen
  \bibfield  {author} {\bibinfo {author} {A.~I. Chirita~Mihaila}, \bibinfo
  {author} {T.~Susi},\ and\ \bibinfo {author} {J.~Kotakoski},\ }\href {\doibase
  10.1038/s41598-019-49565-4} {\bibfield  {journal} {\bibinfo  {journal}
  {Scientific Reports}\ }\textbf {\bibinfo {volume} {9}},\ \bibinfo {pages}
  {12981} (\bibinfo {year} {2019})}\BibitemShut {NoStop}%
\bibitem [{\citenamefont {Susi}\ \emph {\emph{and
  others}}(2017{\natexlab{b}})\citenamefont {Susi}, \citenamefont
  {Kepaptsoglou}, \citenamefont {Lin}, \citenamefont {Ramasse}, \citenamefont
  {Meyer}, \citenamefont {Suenaga},\ and\ \citenamefont
  {Kotakoski}}]{susi_towards_2017}%
  \BibitemOpen
  \bibfield  {author} {\bibinfo {author} {T.~Susi}, \bibinfo {author}
  {D.~Kepaptsoglou}, \bibinfo {author} {Y.-C. Lin}, \bibinfo {author}
  {Q.~Ramasse}, \bibinfo {author} {J.~C. Meyer}, \bibinfo {author}
  {K.~Suenaga},\ and\ \bibinfo {author} {J.~Kotakoski},\ }\href {\doibase
  10.1088/2053-1583/aa878f} {\bibfield  {journal} {\bibinfo  {journal} {2D
  Materials}\ }\textbf {\bibinfo {volume} {4}},\ \bibinfo {pages} {042004}
  (\bibinfo {year} {2017}{\natexlab{b}})}\BibitemShut {NoStop}%
\bibitem [{\citenamefont {Lin}\ \emph {\emph{and others}}(2015)\citenamefont
  {Lin}, \citenamefont {Teng}, \citenamefont {Yeh}, \citenamefont {Koshino},
  \citenamefont {Chiu},\ and\ \citenamefont {Suenaga}}]{lin_structural_2015}%
  \BibitemOpen
  \bibfield  {author} {\bibinfo {author} {Y.-C. Lin}, \bibinfo {author} {P.-Y.
  Teng}, \bibinfo {author} {C.-H. Yeh}, \bibinfo {author} {M.~Koshino},
  \bibinfo {author} {P.-W. Chiu},\ and\ \bibinfo {author} {K.~Suenaga},\ }\href
  {\doibase 10.1021/acs.nanolett.5b02831} {\bibfield  {journal} {\bibinfo
  {journal} {Nano Letters}\ }\textbf {\bibinfo {volume} {15}},\ \bibinfo
  {pages} {7408} (\bibinfo {year} {2015})}\BibitemShut {NoStop}%
\bibitem [{sup(book)}]{supplement}%
  \BibitemOpen
  \href@noop {} {} (\bibinfo {year} {See {S}upplemental {M}aterial at [url],
  containing references
  \cite{romero_abinit_2020,hamann_optimized_2013,setten_pseudodojo_2018,lee_ab_1995,krommer_computational_1998,Mathematica,hourahine_dftb_2020,rauls_stoichiometric_1999},
  for the detailed derivation of the 3{D} energy transfer, the calculated
  phonon density of states and the resulting atomic velocities for the N--C$_2$
  site, full angular variation of its jumping threshold energy, detailed
  description of the theoretical cross section calculation, details on the
  multidimensional numerical integration, including the {W}olfram {M}athematica
  computational notebook.})\BibitemShut {NoStop}%
\bibitem [{\citenamefont {Susi}\ \emph {\emph{and others}}(2012)\citenamefont
  {Susi}, \citenamefont {Kotakoski}, \citenamefont {Arenal}, \citenamefont
  {Kurasch}, \citenamefont {Jiang}, \citenamefont {Skakalova}, \citenamefont
  {Stephan}, \citenamefont {Krasheninnikov}, \citenamefont {Kauppinen},
  \citenamefont {Kaiser},\ and\ \citenamefont {Meyer}}]{susi_atomistic_2012}%
  \BibitemOpen
  \bibfield  {author} {\bibinfo {author} {T.~Susi}, \bibinfo {author}
  {J.~Kotakoski}, \bibinfo {author} {R.~Arenal}, \bibinfo {author}
  {S.~Kurasch}, \bibinfo {author} {H.~Jiang}, \bibinfo {author} {V.~Skakalova},
  \bibinfo {author} {O.~Stephan}, \bibinfo {author} {A.~V. Krasheninnikov},
  \bibinfo {author} {E.~I. Kauppinen}, \bibinfo {author} {U.~Kaiser},\ and\
  \bibinfo {author} {J.~C. Meyer},\ }\href {\doibase 10.1021/nn303944f}
  {\bibfield  {journal} {\bibinfo  {journal} {ACS Nano}\ }\textbf {\bibinfo
  {volume} {6}},\ \bibinfo {pages} {8837} (\bibinfo {year} {2012})}\BibitemShut
  {NoStop}%
\bibitem [{\citenamefont {Larsen}\ \emph {\emph{and others}}(2017)\citenamefont
  {Larsen}, \citenamefont {Mortensen}, \citenamefont {Blomqvist}, \citenamefont
  {Castelli}, \citenamefont {Christensen}, \citenamefont {Dułak},
  \citenamefont {Friis}, \citenamefont {Groves}, \citenamefont {Hammer},
  \citenamefont {Hargus}, \citenamefont {Hermes}, \citenamefont {Jennings},
  \citenamefont {Jensen}, \citenamefont {Kermode}, \citenamefont {Kitchin},
  \citenamefont {Kolsbjerg}, \citenamefont {Kubal}, \citenamefont {Kaasbjerg},
  \citenamefont {Lysgaard}, \citenamefont {Maronsson}, \citenamefont {Maxson},
  \citenamefont {Olsen}, \citenamefont {Pastewka}, \citenamefont {Peterson},
  \citenamefont {Rostgaard}, \citenamefont {Schiøtz}, \citenamefont {Schütt},
  \citenamefont {Strange}, \citenamefont {Thygesen}, \citenamefont {Vegge},
  \citenamefont {Vilhelmsen}, \citenamefont {Walter}, \citenamefont {Zeng},\
  and\ \citenamefont {Jacobsen}}]{larsen_atomic_2017}%
  \BibitemOpen
  \bibfield  {author} {\bibinfo {author} {A.~H. Larsen}, \bibinfo {author}
  {J.~J. Mortensen}, \bibinfo {author} {J.~Blomqvist}, \bibinfo {author} {I.~E.
  Castelli}, \bibinfo {author} {R.~Christensen}, \bibinfo {author} {M.~Dułak},
  \bibinfo {author} {J.~Friis}, \bibinfo {author} {M.~N. Groves}, \bibinfo
  {author} {B.~Hammer}, \bibinfo {author} {C.~Hargus}, \bibinfo {author} {E.~D.
  Hermes}, \bibinfo {author} {P.~C. Jennings}, \bibinfo {author} {P.~B.
  Jensen}, \bibinfo {author} {J.~Kermode}, \bibinfo {author} {J.~R. Kitchin},
  \emph{and others},\ }\href {\doibase 10.1088/1361-648X/aa680e} {\bibfield
  {journal} {\bibinfo  {journal} {Journal of Physics: Condensed Matter}\
  }\textbf {\bibinfo {volume} {29}},\ \bibinfo {pages} {273002} (\bibinfo
  {year} {2017})}\BibitemShut {NoStop}%
\bibitem [{\citenamefont {Enkovaara}\ \emph {\emph{and
  others}}(2010)\citenamefont {Enkovaara}, \citenamefont {Rostgaard},
  \citenamefont {Mortensen}, \citenamefont {Chen}, \citenamefont {Dulak},
  \citenamefont {Ferrighi}, \citenamefont {Gavnholt}, \citenamefont {Glinsvad},
  \citenamefont {Haikola}, \citenamefont {Hansen}, \citenamefont
  {Kristoffersen}, \citenamefont {Kuisma}, \citenamefont {Larsen},
  \citenamefont {Lehtovaara}, \citenamefont {Ljungberg}, \citenamefont
  {Lopez-Acevedo}, \citenamefont {Moses}, \citenamefont {Ojanen}, \citenamefont
  {Olsen}, \citenamefont {Petzold}, \citenamefont {Romero}, \citenamefont
  {Stausholm-Møller}, \citenamefont {Strange}, \citenamefont {Tritsaris},
  \citenamefont {Vanin}, \citenamefont {Walter}, \citenamefont {Hammer},
  \citenamefont {Häkkinen}, \citenamefont {Madsen}, \citenamefont {Nieminen},
  \citenamefont {Nørskov}, \citenamefont {Puska}, \citenamefont {Rantala},
  \citenamefont {Schiøtz}, \citenamefont {Thygesen},\ and\ \citenamefont
  {Jacobsen}}]{enkovaara_electronic_2010}%
  \BibitemOpen
  \bibfield  {author} {\bibinfo {author} {J.~Enkovaara}, \bibinfo {author}
  {C.~Rostgaard}, \bibinfo {author} {J.~J. Mortensen}, \bibinfo {author}
  {J.~Chen}, \bibinfo {author} {M.~Dulak}, \bibinfo {author} {L.~Ferrighi},
  \bibinfo {author} {J.~Gavnholt}, \bibinfo {author} {C.~Glinsvad}, \bibinfo
  {author} {V.~Haikola}, \bibinfo {author} {H.~A. Hansen}, \bibinfo {author}
  {H.~H. Kristoffersen}, \bibinfo {author} {M.~Kuisma}, \bibinfo {author}
  {A.~H. Larsen}, \bibinfo {author} {L.~Lehtovaara}, \bibinfo {author}
  {M.~Ljungberg}, \emph{and others},\ }\href {\doibase
  10.1088/0953-8984/22/25/253202} {\bibfield  {journal} {\bibinfo  {journal}
  {J. Phys. Condens. Matter}\ }\textbf {\bibinfo {volume} {22}},\ \bibinfo
  {pages} {253202} (\bibinfo {year} {2010})}\BibitemShut {NoStop}%
\bibitem [{\citenamefont {Perdew}\ \emph {\emph{and others}}(1996)\citenamefont
  {Perdew}, \citenamefont {Burke},\ and\ \citenamefont
  {Ernzerhof}}]{perdew_generalized_1996}%
  \BibitemOpen
  \bibfield  {author} {\bibinfo {author} {J.~P. Perdew}, \bibinfo {author}
  {K.~Burke},\ and\ \bibinfo {author} {M.~Ernzerhof},\ }\href {\doibase
  10.1103/PhysRevLett.77.3865} {\bibfield  {journal} {\bibinfo  {journal}
  {Phys. Rev. Lett.}\ }\textbf {\bibinfo {volume} {77}},\ \bibinfo {pages}
  {3865} (\bibinfo {year} {1996})}\BibitemShut {NoStop}%
\bibitem [{\citenamefont {Zobelli}\ \emph {\emph{and
  others}}(2007)\citenamefont {Zobelli}, \citenamefont {Gloter}, \citenamefont
  {Ewels}, \citenamefont {Seifert},\ and\ \citenamefont
  {Colliex}}]{zobelli_electron_2007}%
  \BibitemOpen
  \bibfield  {author} {\bibinfo {author} {A.~Zobelli}, \bibinfo {author}
  {A.~Gloter}, \bibinfo {author} {C.~P. Ewels}, \bibinfo {author}
  {G.~Seifert},\ and\ \bibinfo {author} {C.~Colliex},\ }\href {\doibase
  10.1103/PhysRevB.75.245402} {\bibfield  {journal} {\bibinfo  {journal} {Phys.
  Rev. B}\ }\textbf {\bibinfo {volume} {75}},\ \bibinfo {pages} {245402}
  (\bibinfo {year} {2007})}\BibitemShut {NoStop}%
\bibitem [{\citenamefont {Mott}\ and\ \citenamefont
  {Massey}(1965)}]{mott_theory_1965}%
  \BibitemOpen
  \bibfield  {author} {\bibinfo {author} {N.~F. Mott}\ and\ \bibinfo {author}
  {H.~Massey},\ }\href@noop {} {{\selectlanguage {English}\emph {\bibinfo
  {title} {The theory of atomic collisions}}}},\ \bibinfo {edition} {3rd}\ ed.\
  (\bibinfo  {publisher} {Oxford : Clarendon Press},\ \bibinfo {year}
  {1965})\BibitemShut {NoStop}%
\bibitem [{\citenamefont {McKinley}\ and\ \citenamefont
  {Feshbach}(1948)}]{mckinley_coulomb_1948}%
  \BibitemOpen
  \bibfield  {author} {\bibinfo {author} {W.~A. McKinley, Jr.}\ and\ \bibinfo
  {author} {H.~Feshbach},\ }\href {\doibase 10.1103/PhysRev.74.1759} {\bibfield
   {journal} {\bibinfo  {journal} {Phys. Rev.}\ }\textbf {\bibinfo {volume}
  {74}},\ \bibinfo {pages} {1759} (\bibinfo {year} {1948})}\BibitemShut
  {NoStop}%
\bibitem [{\citenamefont {Seitz}\ and\ \citenamefont
  {Koehler}(1956)}]{seitz_notitle_1956}%
  \BibitemOpen
  \bibfield  {author} {\bibinfo {author} {F.~Seitz}\ and\ \bibinfo {author}
  {J.~S. Koehler},\ }in\ \href@noop {} {\emph {\bibinfo {booktitle} {Solid
  {State} {Physics}}}},\ Vol.~\bibinfo {volume} {2}\ (\bibinfo  {publisher}
  {Academic Press},\ \bibinfo {address} {New York},\ \bibinfo {year} {1956})\
  p.\ \bibinfo {pages} {305}\BibitemShut {NoStop}%
\bibitem [{\citenamefont {Kotakoski}\ \emph {\emph{and
  others}}(2010)\citenamefont {Kotakoski}, \citenamefont {Jin}, \citenamefont
  {Lehtinen}, \citenamefont {Suenaga},\ and\ \citenamefont
  {Krasheninnikov}}]{kotakoski_electron_2010}%
  \BibitemOpen
  \bibfield  {author} {\bibinfo {author} {J.~Kotakoski}, \bibinfo {author}
  {C.~Jin}, \bibinfo {author} {O.~Lehtinen}, \bibinfo {author} {K.~Suenaga},\
  and\ \bibinfo {author} {A.~Krasheninnikov},\ }\href {\doibase
  10.1103/PhysRevB.82.113404} {\bibfield  {journal} {\bibinfo  {journal} {Phys.
  Rev. B}\ }\textbf {\bibinfo {volume} {82}},\ \bibinfo {pages} {113404}
  (\bibinfo {year} {2010})}\BibitemShut {NoStop}%
\bibitem [{\citenamefont {Zan}\ \emph {\emph{and others}}(2013)\citenamefont
  {Zan}, \citenamefont {Ramasse}, \citenamefont {Jalil}, \citenamefont
  {Georgiou}, \citenamefont {Bangert},\ and\ \citenamefont
  {Novoselov}}]{zan_control_2013}%
  \BibitemOpen
  \bibfield  {author} {\bibinfo {author} {R.~Zan}, \bibinfo {author} {Q.~M.
  Ramasse}, \bibinfo {author} {R.~Jalil}, \bibinfo {author} {T.~Georgiou},
  \bibinfo {author} {U.~Bangert},\ and\ \bibinfo {author} {K.~S. Novoselov},\
  }\href {\doibase 10.1021/nn4044035} {\bibfield  {journal} {\bibinfo
  {journal} {ACS Nano}\ }\textbf {\bibinfo {volume} {7}},\ \bibinfo {pages}
  {10167} (\bibinfo {year} {2013})}\BibitemShut {NoStop}%
\bibitem [{\citenamefont {Algara-Siller}\ \emph {\emph{and
  others}}(2013)\citenamefont {Algara-Siller}, \citenamefont {Kurasch},
  \citenamefont {Sedighi}, \citenamefont {Lehtinen},\ and\ \citenamefont
  {Kaiser}}]{algara-siller_pristine_2013}%
  \BibitemOpen
  \bibfield  {author} {\bibinfo {author} {G.~Algara-Siller}, \bibinfo {author}
  {S.~Kurasch}, \bibinfo {author} {M.~Sedighi}, \bibinfo {author}
  {O.~Lehtinen},\ and\ \bibinfo {author} {U.~Kaiser},\ }\href {\doibase
  10.1063/1.4830036} {\bibfield  {journal} {\bibinfo  {journal} {Applied
  Physics Letters}\ }\textbf {\bibinfo {volume} {103}},\ \bibinfo {pages}
  {203107} (\bibinfo {year} {2013})}\BibitemShut {NoStop}%
\bibitem [{\citenamefont {Kretschmer}\ \emph {\emph{and
  others}}(2020)\citenamefont {Kretschmer}, \citenamefont {Lehnert},
  \citenamefont {Kaiser},\ and\ \citenamefont
  {Krasheninnikov}}]{kretschmer_formation_2020}%
  \BibitemOpen
  \bibfield  {author} {\bibinfo {author} {S.~Kretschmer}, \bibinfo {author}
  {T.~Lehnert}, \bibinfo {author} {U.~Kaiser},\ and\ \bibinfo {author} {A.~V.
  Krasheninnikov},\ }\href {\doibase 10.1021/acs.nanolett.0c00670} {\bibfield
  {journal} {\bibinfo  {journal} {Nano Letters}\ }\textbf {\bibinfo {volume}
  {20}},\ \bibinfo {pages} {2865} (\bibinfo {year} {2020})}\BibitemShut
  {NoStop}%
\bibitem [{\citenamefont {Lingerfelt}\ \emph {\emph{and
  others}}(2021)\citenamefont {Lingerfelt}, \citenamefont {Yu}, \citenamefont
  {Yoshimura}, \citenamefont {Ganesh}, \citenamefont {Jakowski},\ and\
  \citenamefont {Sumpter}}]{lingerfelt_nonadiabatic_2021}%
  \BibitemOpen
  \bibfield  {author} {\bibinfo {author} {D.~B. Lingerfelt}, \bibinfo {author}
  {T.~Yu}, \bibinfo {author} {A.~Yoshimura}, \bibinfo {author} {P.~Ganesh},
  \bibinfo {author} {J.~Jakowski},\ and\ \bibinfo {author} {B.~G. Sumpter},\
  }\href {\doibase 10.1021/acs.nanolett.0c03587} {\bibfield  {journal}
  {\bibinfo  {journal} {Nano Letters}\ }\textbf {\bibinfo {volume} {21}},\
  \bibinfo {pages} {236} (\bibinfo {year} {2021})}\BibitemShut {NoStop}%
\bibitem [{\citenamefont {Romero}\ \emph {\emph{and others}}(2020)\citenamefont
  {Romero}, \citenamefont {Allan}, \citenamefont {Amadon}, \citenamefont
  {Antonius}, \citenamefont {Applencourt}, \citenamefont {Baguet},
  \citenamefont {Bieder}, \citenamefont {Bottin}, \citenamefont {Bouchet},
  \citenamefont {Bousquet}, \citenamefont {Bruneval}, \citenamefont {Brunin},
  \citenamefont {Caliste}, \citenamefont {Côté}, \citenamefont {Denier},
  \citenamefont {Dreyer}, \citenamefont {Ghosez}, \citenamefont {Giantomassi},
  \citenamefont {Gillet}, \citenamefont {Gingras}, \citenamefont {Hamann},
  \citenamefont {Hautier}, \citenamefont {Jollet}, \citenamefont {Jomard},
  \citenamefont {Martin}, \citenamefont {Miranda}, \citenamefont {Naccarato},
  \citenamefont {Petretto}, \citenamefont {Pike}, \citenamefont {Planes},
  \citenamefont {Prokhorenko}, \citenamefont {Rangel}, \citenamefont {Ricci},
  \citenamefont {Rignanese}, \citenamefont {Royo}, \citenamefont {Stengel},
  \citenamefont {Torrent}, \citenamefont {van Setten}, \citenamefont
  {Van~Troeye}, \citenamefont {Verstraete}, \citenamefont {Wiktor},
  \citenamefont {Zwanziger},\ and\ \citenamefont {Gonze}}]{romero_abinit_2020}%
  \BibitemOpen
  \bibfield  {author} {\bibinfo {author} {A.~H. Romero}, \bibinfo {author}
  {D.~C. Allan}, \bibinfo {author} {B.~Amadon}, \bibinfo {author}
  {G.~Antonius}, \bibinfo {author} {T.~Applencourt}, \bibinfo {author}
  {L.~Baguet}, \bibinfo {author} {J.~Bieder}, \bibinfo {author} {F.~Bottin},
  \bibinfo {author} {J.~Bouchet}, \bibinfo {author} {E.~Bousquet}, \bibinfo
  {author} {F.~Bruneval}, \bibinfo {author} {G.~Brunin}, \bibinfo {author}
  {D.~Caliste}, \bibinfo {author} {M.~Côté}, \bibinfo {author} {J.~Denier},
  \emph{and others},\ }\href {\doibase 10.1063/1.5144261} {\bibfield  {journal}
  {\bibinfo  {journal} {The Journal of Chemical Physics}\ }\textbf {\bibinfo
  {volume} {152}},\ \bibinfo {pages} {124102} (\bibinfo {year}
  {2020})}\BibitemShut {NoStop}%
\bibitem [{\citenamefont {Hamann}(2013)}]{hamann_optimized_2013}%
  \BibitemOpen
  \bibfield  {author} {\bibinfo {author} {D.~R. Hamann},\ }\href {\doibase
  10.1103/PhysRevB.88.085117} {\bibfield  {journal} {\bibinfo  {journal} {Phys.
  Rev. B}\ }\textbf {\bibinfo {volume} {88}},\ \bibinfo {pages} {085117}
  (\bibinfo {year} {2013})}\BibitemShut {NoStop}%
\bibitem [{\citenamefont {Setten}\ \emph {\emph{and others}}(2018)\citenamefont
  {Setten}, \citenamefont {Giantomassi}, \citenamefont {Bousquet},
  \citenamefont {Verstraete}, \citenamefont {Hamann}, \citenamefont {Gonze},\
  and\ \citenamefont {Rignanese}}]{setten_pseudodojo_2018}%
  \BibitemOpen
  \bibfield  {author} {\bibinfo {author} {M.~J.~v. Setten}, \bibinfo {author}
  {M.~Giantomassi}, \bibinfo {author} {E.~Bousquet}, \bibinfo {author} {M.~J.
  Verstraete}, \bibinfo {author} {D.~R. Hamann}, \bibinfo {author} {X.~Gonze},\
  and\ \bibinfo {author} {G.-M. Rignanese},\ }\href@noop {} {\bibfield
  {journal} {\bibinfo  {journal} {Computer Physics Communications}\ }\textbf
  {\bibinfo {volume} {226}},\ \bibinfo {pages} {39} (\bibinfo {year}
  {2018})}\BibitemShut {NoStop}%
\bibitem [{\citenamefont {Lee}\ and\ \citenamefont
  {Gonze}(1995)}]{lee_ab_1995}%
  \BibitemOpen
  \bibfield  {author} {\bibinfo {author} {C.~Lee}\ and\ \bibinfo {author}
  {X.~Gonze},\ }\href {\doibase 10.1103/PhysRevB.51.8610} {\bibfield  {journal}
  {\bibinfo  {journal} {Phys. Rev. B}\ }\textbf {\bibinfo {volume} {51}},\
  \bibinfo {pages} {8610} (\bibinfo {year} {1995})}\BibitemShut {NoStop}%
\bibitem [{\citenamefont {Krommer}\ and\ \citenamefont
  {Ueberhuber}(1998)}]{krommer_computational_1998}%
  \BibitemOpen
  \bibfield  {author} {\bibinfo {author} {A.~R. Krommer}\ and\ \bibinfo
  {author} {C.~W. Ueberhuber},\ }\href {10.1137/1.9781611971460} {\emph
  {\bibinfo {title} {Computational integration}}}\ (\bibinfo  {publisher}
  {SIAM},\ \bibinfo {year} {1998})\BibitemShut {NoStop}%
\bibitem [{\citenamefont {Inc.}(2021)}]{Mathematica}%
  \BibitemOpen
  \bibfield  {author} {\bibinfo {author} {W.~R. Inc.},\ }\href
  {https://www.wolfram.com/mathematica} {\enquote {\bibinfo {title}
  {Mathematica, {V}ersion 12.3.1},}\ } (\bibinfo {year} {2021})\BibitemShut
  {NoStop}%
\bibitem [{\citenamefont {Hourahine}\ \emph {\emph{and
  others}}(2020)\citenamefont {Hourahine}, \citenamefont {Aradi}, \citenamefont
  {Blum}, \citenamefont {Bonafé}, \citenamefont {Buccheri}, \citenamefont
  {Camacho}, \citenamefont {Cevallos}, \citenamefont {Deshaye}, \citenamefont
  {Dumitrică}, \citenamefont {Dominguez}, \citenamefont {Ehlert},
  \citenamefont {Elstner}, \citenamefont {van~der Heide}, \citenamefont
  {Hermann}, \citenamefont {Irle}, \citenamefont {Kranz}, \citenamefont
  {Köhler}, \citenamefont {Kowalczyk}, \citenamefont {Kubař}, \citenamefont
  {Lee}, \citenamefont {Lutsker}, \citenamefont {Maurer}, \citenamefont {Min},
  \citenamefont {Mitchell}, \citenamefont {Negre}, \citenamefont {Niehaus},
  \citenamefont {Niklasson}, \citenamefont {Page}, \citenamefont {Pecchia},
  \citenamefont {Penazzi}, \citenamefont {Persson}, \citenamefont {Řezáč},
  \citenamefont {Sánchez}, \citenamefont {Sternberg}, \citenamefont {Stöhr},
  \citenamefont {Stuckenberg}, \citenamefont {Tkatchenko}, \citenamefont {Yu},\
  and\ \citenamefont {Frauenheim}}]{hourahine_dftb_2020}%
  \BibitemOpen
  \bibfield  {author} {\bibinfo {author} {B.~Hourahine}, \bibinfo {author}
  {B.~Aradi}, \bibinfo {author} {V.~Blum}, \bibinfo {author} {F.~Bonafé},
  \bibinfo {author} {A.~Buccheri}, \bibinfo {author} {C.~Camacho}, \bibinfo
  {author} {C.~Cevallos}, \bibinfo {author} {M.~Y. Deshaye}, \bibinfo {author}
  {T.~Dumitrică}, \bibinfo {author} {A.~Dominguez}, \bibinfo {author}
  {S.~Ehlert}, \bibinfo {author} {M.~Elstner}, \bibinfo {author} {T.~van~der
  Heide}, \bibinfo {author} {J.~Hermann}, \bibinfo {author} {S.~Irle},
  \emph{and others},\ }\href {\doibase 10.1063/1.5143190} {\bibfield  {journal}
  {\bibinfo  {journal} {The Journal of Chemical Physics}\ }\textbf {\bibinfo
  {volume} {152}},\ \bibinfo {pages} {124101} (\bibinfo {year}
  {2020})}\BibitemShut {NoStop}%
\bibitem [{\citenamefont {Rauls}\ \emph {\emph{and others}}(1999)\citenamefont
  {Rauls}, \citenamefont {Elsner}, \citenamefont {Gutierrez},\ and\
  \citenamefont {Frauenheim}}]{rauls_stoichiometric_1999}%
  \BibitemOpen
  \bibfield  {author} {\bibinfo {author} {E.~Rauls}, \bibinfo {author}
  {J.~Elsner}, \bibinfo {author} {R.~Gutierrez},\ and\ \bibinfo {author}
  {T.~Frauenheim},\ }\href {\doibase
  https://doi.org/10.1016/S0038-1098(99)00137-4} {\bibfield  {journal}
  {\bibinfo  {journal} {Solid State Communications}\ }\textbf {\bibinfo
  {volume} {111}},\ \bibinfo {pages} {459} (\bibinfo {year}
  {1999})}\BibitemShut {NoStop}%
\end{thebibliography}
\end{document}